\documentclass[preprint2]{aastex}

\usepackage{psfig}
\usepackage{lscape}
\usepackage{natbib}

\def\cte[#1]{$^{#1}$}

\shorttitle{Sub-Milliarcsecond Imaging of AGN}
\shortauthors{J.A. Zensus et al.}

\begin{document}

\title{Sub-Milliarcsecond Imaging of Quasars and Active Galactic Nuclei\\
    II. Additional Sources}

\author{J. A. Zensus}
\affil{Max-Planck-Institut f\"ur Radioastronomie, Auf dem H\"ugel 69, D-53121 Bonn, 
Germany \and
National Radio Astronomy Observatory, 520 Edgemont Rd., Charlottesville, VA 22903}
\email{azensus@mpifr-bonn.mpg.de}

\author{E. Ros}
\affil{Max-Planck-Institut f\"ur Radioastronomie, Auf dem H\"ugel 69, 
D-53121 Bonn, Germany}
\email{ros@mpifr-bonn.mpg.de}

\author{K. I. Kellermann}
\affil{National Radio Astronomy Observatory, 520 Edgemont Rd., Charlottesville, VA 22903}
\email{kkellerm@nrao.edu}

\author{M. H. Cohen}
\affil{California Institute of Technology, Department of Astronomy, MS 105-24, 
Pasadena, CA 91125}
\email{mhc@astro.caltech.edu}

\author{R. C. Vermeulen}
\affil{Netherlands Foundation for Research in Astronomy, Postbus 2, 7990\,AA
Dwingeloo, The Netherlands}
\email{rvermeulen@nfra.nl}

\author{M. Kadler}
\affil{Max-Planck-Institut f\"ur Radioastronomie, Auf dem H\"ugel 69, 
D-53121 Bonn, Germany}
\email{mkadler@mpifr-bonn.mpg.de}
\date{Submitted: Feb 22 2002 - Accepted: Apr 30 2002}

\begin{abstract}
We report further results from our imaging survey at 15\,GHz ($\lambda$=2\,cm)
with the Very Long Baseline
Array. This paper presents single epoch images for 39 sources,
bringing the total number of objects in the sample to 171.  Our sample is
representative of a complete unbiased sample and it 
will be used for statistical
analysis of source properties. We compare the observed brightness
temperatures derived from our VLBA observations to those
derived from total intensity variations at 22 and 37\,GHz.  These
are consistent with intrinsic brightness temperatures in the range
$10^{10}$ to $10^{12}$\,K.
We also present three new spectroscopic redshift values: 
$z$=0.517$\pm0.001$ for 0026+346,
$z$=1.591$\pm0.003$ for 0727--115, 
and $z$=0.2016$\pm0.0004$ for 1155+251.

Images from this VLBA 2\,cm survey are available on
the Internet under \url{http://www.cv.nrao.edu/2cmsurvey}.
\end{abstract}


\section{Introduction}

We have been using the Very Long Baseline Array\footnote{The National Radio
Astronomy Observatory and the Very Long Baseline Array are operated by
Associated Universities, Inc., under cooperative agreement with the
U.S.\ National Science Foundation} (VLBA; \citealt{Napier94}) since 1994 to
study a sample of compact radio sources at 15\,GHz ($\lambda = 2$~cm). We are
particularly interested in the scientific aspects related to: a) the
morphology of the compact radio sources associated with quasars and radio
galaxies (see \citealt{Zensus97}); 
this includes comparisons with results from
longer-wavelength studies; b) source variability; c) statistics of internal
kinematics; d) correlations of the radio results with other properties, e.g.,
redshift and high-energy luminosity. Our ultimate goal is to derive
implications for the
relationship between different classes of AGN, for their physical and
cosmological evolution, and for the physics of radio jets.

The images and results of the first 132 sources of our sample
were published by
\citet{kellermann98} (hereafter Paper I). 
The ``complete'' sample 
includes 39 additional sources. Images of these objects and three new
redshifts are presented and discussed in this paper. The results of
the full project from all epochs are accessible on the Internet at
\url{http://www.cv.nrao.edu/2cmsurvey}.

Throughout the paper we use a Hubble constant,
$H_0$=65\,km\,s$^{-1}$\,Mpc$^{-1}$ and the deceleration parameter,
$q_0$=0.5.

\section{Definition of the sample \label{sec:sampledefinition}}

Our sample consists of 171 sources that were observed
during the period 1994-2002.  Our initial selection criterion included all
sources from the catalog of \citet{Stickel94} which are strong (S$_{\rm
15\,GHz}>$1.5\,Jy for $\delta>0^\circ$ and S$_{\rm 15\,GHz}>$2\,Jy for
$0^\circ>\delta>-20^\circ$), and which have a flat radio spectrum 
($\alpha>-0.5$, S$\sim\nu^{+\alpha}$) at any frequency above 
500\,MHz. The Stickel catalog
is complete only at 5\,GHz, and we have
used other measurements or
extrapolations to form our 15\,GHz sample (see Paper I).  We have
now observed 93 of the 124 sources that fit the
above
criteria. The remaining 31 sources\footnote{The missing sources are:
0113--118$^*$, 0138--097, 0146+056$^*$, 0248+430$^*$, 0256+075,
0332+078, 0400+258, 0403--102$^*$, 0446+112$^*$, 0539--057,
0743--006, 0831+557$^*$, 0833+585, 0954+556, 0954+658,
1030+415, 1039+811, 1147+245, 1216+487, 1222+131,
1418+546, 1637+574, 1725+044, 1732+389, 1751+441,
1936--155$^*$, 2008--159$^*$, 2029+121$^*$, 2203--188$^*$, 2216--038, 2254+074.
An asterisk indicates that this source was observed in as 
part of a separate
program by Gurvits, Fomalont, \& Kellermann (in preparation).
}
were not observed for a variety of logistic and technical reasons
(especially scheduling constraints due to limited observing
time), however we consider the 
93~sources as representative of an unbiased sample.

An additional 78~sources
did not originally fit our criterion but were added because subsequent
measurements, including RATAN\,600 observations at 15\,GHz
(\citealt{Kovalev99}), showed that they did.  They are not in the
Stickel catalogue.

Most, if not all, flat-spectrum sources are variable. Thus, at any given time,
some sources with mean flux density below the flux limit of a survey will
nevertheless be included because they are in a high state, and others
with mean flux density above the flux limit will not make it into the sample
because they happen to be in a low state. Because of the steep source
count or $N(S)$ function, there will be more sources ``inappropriately''
included than are missed. To the extent that the flux density that we used to
determine our sample is from a single measurement, the statistical content of
the sample is independent of the epochs of measurement; i.e.\ of the specific
sources which comprise the sample (see also \citealt{Drinkwater97}).
In our case we have a mixture of measurements at 15\,GHz and some
extrapolations from lower frequencies.  Although it is not possible to
quantify the statistical effect of this mixture, we conclude
that our list is
representative of an unbiased,
flux-limited sample. As such it is useful for statistical
studies of strong compact radio sources.

\citet{Pearson88} have presented a complete sample of 65 sources defined by
$\delta>$35$^\circ$; $\left|b\right|>$10$^\circ$; and S$_{\rm
5GHz}\geq$1.3\,Jy.  32 of these do not match either our flux density or our
spectral index criterion. We have observed 31 of the remaining 33 objects.
Thus there are only two sources\footnote{0831+557, 0954+556}
missing from the complete sample defined by both
the selection criteria of the \citet{Pearson88}
sample and ours.

New spectroscopic redshifts for three sources are reported here:
0026+346 with $z$=0.517$\pm0.001$,
0727--115 with $z$=1.591$\pm0.003$, 
and 1155+251 with $z$=0.2016$\pm0.0004$; details are given in Appendix
\ref{app:redshift}. This now leaves 9 sources out
of the 171 without a redshift.  
Two of these sources are in empty fields (i.e., regions without
compact optical features brighter than 20\,mag), and
identifications for the others are shown in Table~\ref{table:sourcelist}. These
identifications are taken from the literature with the aid of the NASA/IPAC
Extragalactic Database (NED) and the SIMBAD Database.
The radio galaxies are optical galaxies identified with sources in radio 
catalogs.  Quasars are optically unresolved objects with broad emission
lines and, in some cases, absorption lines.  BL\,Lac objects have weak or
no emission lines in their spectra.

\begin{deluxetable}{lp{45mm}cccrlrc}
\tabletypesize{\scriptsize}
\tablecaption{Source List. \label{table:sourcelist}}
\tablehead{
\colhead{} &
\colhead{} &
\colhead{R.A.} &
\colhead{Decl.} &
\colhead{} &
\colhead{} &
\colhead{} &
\colhead{$S_6$\tablenotemark{g}} &
\colhead{\# of}\\
\colhead{Source\tablenotemark{a}} &
\colhead{Name\tablenotemark{b}} &
\colhead{(J2000.0)\tablenotemark{c}} &
\colhead{(J2000.0)\tablenotemark{c}} &
\colhead{ID\tablenotemark{d}} &
\colhead{$V$\tablenotemark{e}} &
\colhead{$z$\tablenotemark{f}} &
\colhead{(Jy)} &
\colhead{epochs}
}
\startdata
\multicolumn{9}{c}{Sources presented in this paper} \\ \hline
0108$+$388&                         & $01^{\rm h}11^{\rm m}37\rlap{.}^{\rm s}39$ & $+39^\circ06^\prime28\rlap{.}^{\prime\prime}10$ & G & 22.00 & 0.668$^{(1)}$ & 1.34 & 1 \\
0119$+$115&                         & $01^{\rm h}21^{\rm m}41\rlap{.}^{\rm s}67$ & $+11^\circ49^\prime50\rlap{.}^{\prime\prime}60$ & Q & 19.50 & 0.57$^{(1)}$  & 1.01 & 1 \\
0201$+$113&                         & $02^{\rm h}03^{\rm m}46\rlap{.}^{\rm s}72$ & $+11^\circ34^\prime45\rlap{.}^{\prime\prime}60$ & Q & 19.50 & 3.56$^{(1)}$  & 0.84 & 1 \\
0221$+$067& 4C\,06.11               & $02^{\rm h}24^{\rm m}28\rlap{.}^{\rm s}49$ & $+06^\circ59^\prime23\rlap{.}^{\prime\prime}50$ & Q & 20.00 & 0.511$^{(1)}$ & 1.03 & 2 \\ 
0310$+$013&                         & $03^{\rm h}12^{\rm m}43\rlap{.}^{\rm s}60$ & $+01^\circ33^\prime17\rlap{.}^{\prime\prime}54$ & Q & 18.24 & 0.664$^{(2)}$ & 0.72 & 1 \\ 
0405$-$385&                         & $04^{\rm h}06^{\rm m}59\rlap{.}^{\rm s}07$ & $-38^\circ26^\prime27\rlap{.}^{\prime\prime}80$ & Q & 18.00 & 1.285$^{(3)}$ & 1.09 & 1 \\ 
0420$+$022&                         & $04^{\rm h}22^{\rm m}52\rlap{.}^{\rm s}22$ & $+02^\circ19^\prime26\rlap{.}^{\prime\prime}94$ & BL& 19.50 &  ...          & 1.22 & 1  \\ 
0723$-$008&                         & $07^{\rm h}25^{\rm m}50\rlap{.}^{\rm s}69$ & $-00^\circ54^\prime56\rlap{.}^{\prime\prime}60$ & BL& 17.50 & 0.127$^{(1)}$&  1.11\tablenotemark{i} & 1 \\ 
0834$-$201&             	    & $08^{\rm h}36^{\rm m}39\rlap{.}^{\rm s}26$ & $-20^\circ16^\prime59\rlap{.}^{\prime\prime}70$ & Q & 19.40 & 2.752$^{(5)}$& 3.72 & 2 \\ 
0836$+$710\tablenotemark{h}
          & 4C\,71.07		    & $08^{\rm h}41^{\rm m}24\rlap{.}^{\rm s}41$ & $+70^\circ53^\prime42\rlap{.}^{\prime\prime}20$ & Q & 16.50 & 2.172$^{(6)}$& 2.59 & 1 \\ 
0859$+$470& 4C\,47.29               & $09^{\rm h}03^{\rm m}04\rlap{.}^{\rm s}05$ & $+46^\circ51^\prime04\rlap{.}^{\prime\prime}10$ & Q & 18.70 & 1.462$^{(7)}$& 1.78 & 1 \\ 
0906$+$015& 4C\,01.24               & $09^{\rm h}09^{\rm m}10\rlap{.}^{\rm s}15$ & $+01^\circ21^\prime35\rlap{.}^{\prime\prime}50$ & Q & 17.50 & 1.018$^{(1)}$& 1.04 & 4 \\ 
1032$-$199&                         & $10^{\rm h}35^{\rm m}02\rlap{.}^{\rm s}21$ & $-20^\circ11^\prime34\rlap{.}^{\prime\prime}70$ & Q & 19.00 & 2.198$^{(1)}$& 1.15 & 1 \\ 
1124$-$186&          		    & $11^{\rm h}27^{\rm m}04\rlap{.}^{\rm s}46$ & $-18^\circ57^\prime17\rlap{.}^{\prime\prime}80$ & Q & 19.00 & 1.048$^{(1)}$& 1.62 & 1 \\ 
1145$-$071&                         & $11^{\rm h}47^{\rm m}51\rlap{.}^{\rm s}62$ & $-07^\circ24^\prime41\rlap{.}^{\prime\prime}40$ & Q & 18.50 & 1.342$^{(1)}$& 1.25 & 1 \\ 
1148$-$001& 4C\,-00.47	            & $11^{\rm h}50^{\rm m}43\rlap{.}^{\rm s}87$ & $-00^\circ23^\prime54\rlap{.}^{\prime\prime}20$ & Q & 17.60 & 1.980$^{(1)}$& 1.90 & 1 \\ 
1345$+$125&          4C\,12.50	    & $13^{\rm h}47^{\rm m}33\rlap{.}^{\rm s}44$ & $+12^\circ17^\prime24\rlap{.}^{\prime\prime}10$ & G & 17.00 & 0.121$^{(1)}$& 2.91 & 3 \\ 
1354$-$152&               	    & $13^{\rm h}57^{\rm m}11\rlap{.}^{\rm s}33$ & $-15^\circ27^\prime29\rlap{.}^{\prime\prime}00$ & Q & 18.50 & 1.89$^{(1)}$ & 1.52 & 2 \\ 
1354$+$196& 4C\,19.44, DA\,354   
                                    & $13^{\rm h}57^{\rm m}04\rlap{.}^{\rm s}51$ & $+19^\circ19^\prime07\rlap{.}^{\prime\prime}30$ & Q & 16.02 & 0.719$^{(1)}$ & 1.56 & 2 \\ 
1458$+$718& 4C\,71.15, 3C\,309.1, NRAO\,0464
                                    & $14^{\rm h}59^{\rm m}07\rlap{.}^{\rm s}66$ & $+71^\circ40^\prime20\rlap{.}^{\prime\prime}00$ & Q & 16.78 & 0.905$^{(8)}$& 3.39 & 2 \\ 
1502$+$106&  4C\,10.39	            & $15^{\rm h}04^{\rm m}25\rlap{.}^{\rm s}06$ & $+10^\circ29^\prime39\rlap{.}^{\prime\prime}10$ & Q & 15.50 & 1.833$^{(1)}$& 2.56 & 2 \\ 
1504$-$167\tablenotemark{h}
          &                         & $15^{\rm h}07^{\rm m}04\rlap{.}^{\rm s}88$ & $-16^\circ52^\prime30\rlap{.}^{\prime\prime}50$ & Q & 18.50 & 0.876$^{(1)}$& 1.98 & 2 \\ 
1504$+$377&       		    & $15^{\rm h}06^{\rm m}09\rlap{.}^{\rm s}61$ & $+37^\circ30^\prime51\rlap{.}^{\prime\prime}20$ & G & 21.20 & 0.674$^{(1)}$& 1.10 & 1 \\ 
1511$-$100&          		    & $15^{\rm h}13^{\rm m}44\rlap{.}^{\rm s}98$ & $+10^\circ12^\prime00\rlap{.}^{\prime\prime}40$ & Q & 18.50 & 1.513$^{(1)}$& 1.22 & 1 \\ 
1514$-$241& AP\,Libr{\ae}           & $15^{\rm h}17^{\rm m}41\rlap{.}^{\rm s}91$ & $-24^\circ22^\prime19\rlap{.}^{\prime\prime}70$ & BL& 15.00 & 0.0486$^{(1)}$& 2.00 & 1 \\ 
1519$-$273&                         & $15^{\rm h}22^{\rm m}37\rlap{.}^{\rm s}77$ & $-27^\circ30^\prime11\rlap{.}^{\prime\prime}00$ & BL & 18.50 & 0.071        & 2.35 & 1 \\ 
1538$+$149&           4C\,14.6, 2E\,3486  
                                    & $15^{\rm h}40^{\rm m}49\rlap{.}^{\rm s}58$ & $+14^\circ47^\prime45\rlap{.}^{\prime\prime}80$ & BL & 15.50 & 0.605$^{(1)}$& 1.96 & 1  \\ 
1555$+$001& DA\,393                 & $15^{\rm h}57^{\rm m}51\rlap{.}^{\rm s}52$ & $-00^\circ01^\prime50\rlap{.}^{\prime\prime}50$ & Q & 19.30 & 1.772$^{(1)}$& 2.24  & 2 \\ 
1622$-$253\tablenotemark{h}
         & ROS\,1	            & $16^{\rm h}25^{\rm m}46\rlap{.}^{\rm s}89$ & $-25^\circ27^\prime38\rlap{.}^{\prime\prime}30$ & Q & 20.60 & 0.786$^{(1)}$& 2.08 & 2 \\ 
1739$+$522\tablenotemark{h}
         & 4C\,51.37, 2E\,3936
                                    & $17^{\rm h}40^{\rm m}36\rlap{.}^{\rm s}07$ & $+52^\circ11^\prime43\rlap{.}^{\prime\prime}50$ & Q & 18.50 & 1.375$^{(6)}$& 1.98 & 1 \\ 
1828$+$487& 3C\,380, 4C\,48.46, NRAO\,565, CTA\,79
                                    & $18^{\rm h}29^{\rm m}31\rlap{.}^{\rm s}80$ & $+48^\circ44^\prime46\rlap{.}^{\prime\prime}62$ & Q & 16.81 & 0.692$^{(7)}$& 6.19 & 2 \\ 
1954$+$513&        		    & $19^{\rm h}55^{\rm m}42\rlap{.}^{\rm s}83$ & $+51^\circ31^\prime48\rlap{.}^{\prime\prime}60$ & Q & 18.50 & 1.223$^{(9)}$ & 1.43 & 2 \\ 
2121$+$053&             	    & $21^{\rm h}23^{\rm m}44\rlap{.}^{\rm s}61$ & $+05^\circ35^\prime22\rlap{.}^{\prime\prime}30$ & Q & 17.50 & 1.941$^{(1)}$& 1.18\tablenotemark{i} & 2  \\ 
2128$-$123&                       & $21^{\rm h}31^{\rm m}35\rlap{.}^{\rm s}35$ & $-12^\circ07^\prime04\rlap{.}^{\prime\prime}50$ & Q & 15.98 & 0.501$^{(1)}$& 2.07 & 3 \\ 
2128$+$048&                       & $21^{\rm h}31^{\rm m}35\rlap{.}^{\rm s}35$ & $-12^\circ07^\prime04\rlap{.}^{\prime\prime}50$ & Q & 15.98 & 0.501$^{(1)}$& 2.07 & 1 \\ 
2155$-$152&             	    & $21^{\rm h}58^{\rm m}06\rlap{.}^{\rm s}37$ & $-15^\circ01^\prime09\rlap{.}^{\prime\prime}00$ & Q & 17.50 & 0.672$^{(1)}$& 1.77 & 1 \\ 
2227$-$088& PHL\,5225		    & $22^{\rm h}29^{\rm m}40\rlap{.}^{\rm s}17$ & $-08^\circ32^\prime54\rlap{.}^{\prime\prime}10$ & Q & 17.50 & 1.562$^{(1)}$& 1.77 & 1 \\ 
2255$-$282\tablenotemark{h}
         & 			    & $22^{\rm h}58^{\rm m}06\rlap{.}^{\rm s}05$ & $-27^\circ58^\prime20\rlap{.}^{\prime\prime}90$ & Q & 16.77 & 0.926$^{(1)}$& 1.78 & 1 \\ 
2318$+$049&                          & $23^{\rm h}20^{\rm m}44\rlap{.}^{\rm s}94$ & $+05^\circ13^\prime50\rlap{.}^{\prime\prime}20$ & Q & 19.00 & 0.623$^{(1)}$& 1.13 & 2 \\  \hline 
\multicolumn{9}{c}{Sources from Table 1 in Paper I with newer parameters} \\ \hline
0026$+$346 & & $00^{\rm h}29^{\rm m}14\rlap{.}^{\rm s}24$ & $+34^\circ56^\prime32\rlap{.}^{\prime\prime}26$ & G   & 20.17 & 0.517$^{(10)}$ & 1.27 &  4 \\
0202$+$149 & & $02^{\rm h}04^{\rm m}50\rlap{.}^{\rm s}41$ & $+15^\circ14^\prime11\rlap{.}^{\prime\prime}04$ & G  & 22.10 & 0.405$^{(11)}$ & 2.47 & 5 \\
0727$-$115 & & $07^{\rm h}30^{\rm m}19\rlap{.}^{\rm s}11$ & $-11^\circ41^\prime12\rlap{.}^{\prime\prime}60$ & G? & 20.30   & 1.591$^{(10)}$  & 2.20 & 5 \\
0814$+$425 & & $08^{\rm h}18^{\rm m}16\rlap{.}^{\rm s}00$ & $+42^\circ22^\prime45\rlap{.}^{\prime\prime}41$ & BL  & 18.50 & 0.245$^{(12)}$ & 1.69 & 7 \\
1155$+$251 & & $11^{\rm h}58^{\rm m}25\rlap{.}^{\rm s}79$ & $+24^\circ50^\prime18\rlap{.}^{\prime\prime}00$ & G  & 17.50 & 0.202$^{(10)}$ & 1.16 & 5 \\
1228$+$126 & M87 & $12^{\rm h}30^{\rm m}49\rlap{.}^{\rm s}42$ & $+12^\circ23^\prime28\rlap{.}^{\prime\prime}04$ & G   & 9.60  & 0.0044$^{(13)}$ & 74.90 & 11 \\
1302$-$102 & & $13^{\rm h}05^{\rm m}33\rlap{.}^{\rm s}02$ & $-10^\circ33^\prime19\rlap{.}^{\prime\prime}43$ & Q   & 14.90 & 0.278$^{(14)}$ & 1.17 & 5 \\
1413$+$135 & & $14^{\rm h}15^{\rm m}58\rlap{.}^{\rm s}82$ & $+13^\circ20^\prime23\rlap{.}^{\prime\prime}71$ & BL  & 20.00 & 0.247$^{(15)}$ & 0.85 & 6 \\
1656$+$053 & & $16^{\rm h}58^{\rm m}33\rlap{.}^{\rm s}45$ & $+05^\circ15^\prime16\rlap{.}^{\prime\prime}44$ & Q   & 16.50 & 0.879$^{(16)}$ & 2.16 & 4 \\
2131$-$021 & & $21^{\rm h}34^{\rm m}10\rlap{.}^{\rm s}31$ & $-01^\circ53^\prime17\rlap{.}^{\prime\prime}24$ & BL  & 18.70 & 1.285$^{(17)}$ & 2.12 & 4 \\
\tableline
\enddata

\tablenotetext{a}{ IAU source designation.}
\tablenotetext{b}{ Alternative source name where appropriate.}
\tablenotetext{c}{ Right ascension and declination (J2000.0).}
\tablenotetext{d}{ The optical counterpart, denoted as follows: (G) galaxy,
(Q) quasar, or (BL) BL\,Lac object 
}
\tablenotetext{e}{ Optical magnitude.}
\tablenotetext{f}{ Redshift.}
\tablenotetext{g}{ Flux density at 5\,GHz, taken mostly from
\citet{Stickel94}.}
\tablenotetext{h}{Detected by the EGRET instrument on the Compton Gamma Ray Observatory
(\citealt{mat97,Macomb99,Hartman99}).}
\tablenotetext{i}{Taken from the UMRAO Radio Survey Flux Database 
(http://www.astro.lsa.umich.edu/obs/radiotel/umrao.html, see 
\citealt{hughes92}).}
\tablerefs{
(1) \citet{Kovalev99} 
(2) \citet{Hewett95} 
(3) \citet{Stickel94} 
(4) \citet{Keel85} 
(5) \citet{Hewitt89} 
(6) \citet{hutsemekers98} 
(7) \citet{machalski98} 
(8) \citet{nilsson98} 
(9) \citet{Polatidis95} 
(10) See Appendix 
(11) \citet{Perlman98} 
(12) \citet{Lawrence96} 
(13) \citet{Smith00} 
(14) \citet{Marziani96} 
(15) \citet{Wiklind97} 
(16) \citet{Pravdo84} 
(17) \citet{Drinkwater97} 
}
\end{deluxetable}

\begin{figure}[htbp]
\figurenum{1}
{\psfig{figure=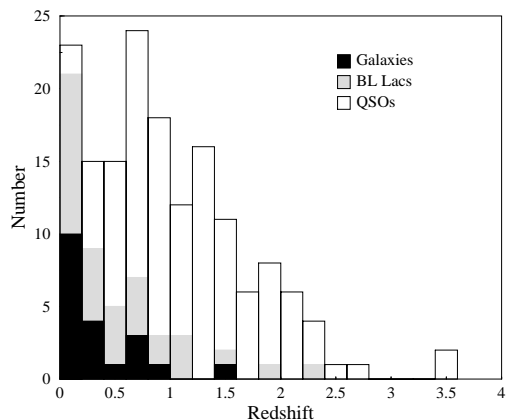,width=0.40\textwidth,angle=-90}}
\caption{Distribution of redshifts for our 171 sources. 
\label{fig:hist}
}
\end{figure}

Figure~\ref{fig:hist} shows the distribution of redshifts for the
sample. The galaxies and
BL\,Lacs are concentrated at low redshift, while the quasars have a broad peak
around $z$=1. This is similar to the redshift distribution of the total
\citet{Stickel94} sample; see their Figure~1.

\section{Observations \label{sec:observations}}

Table~\ref{table:sourcelist} lists the 39 additional sources and, at bottom,
the 10 sources from Table~1 of Paper I for which new
information has become available such as the redshift or identification.
We use these new data in the discussion below.

We have continued our observing program described in Paper I with 12
additional epochs, from August 1997 to March 2001.
Throughout, we observed with a bandwidth of 64\,MHz using 1-bit samples and
left-circular polarization.  Each source was observed for 4--6 minutes once
per hour, over a range of 8\,hours.  Three groups were observed
per day with a maximum of 30\,sources per day.  Typical integration times
on each source were 48\,min.

The amplitude and phase calibration of the visibility
data across the frequency channels and across time (usually refered to
as {\it a priori}
calibration and fringe fitting) were 
carried out for each
source using techniques contained
within {\sc aips}.
Initial hybrid imaging iterations and self-calibration proceeded
with a standard {\sc difmap} script (\citealt{Pearson94}); 
hands-on further processing was carried out whenever
obvious problems arose. More details on our data reduction methods are given
in Paper I.

\section{Discussion \label{sec:discussion}}

Table~\ref{table:sourcestructure} lists the source properties derived from the
contour diagrams which are shown in Figure~\ref{fig:maps}. The columns are,
successively, the IAU name, the total flux density at 15\,GHz calculated from
the VLBA images, the luminosity, the epoch, the lowest contour level
shown in Figure~\ref{fig:maps} (equal to 3~times the root-mean-square noise),
the major axis, minor axis, and position angle of the restoring beam, 
the peak brightness temperature from the images, and the 
morphological classification.
The equation describing the brightness temperature given in Paper I
contains two errors, although the entries in Table 3 (col 10) are
correct.  The correct expression which we have used in calculating the
values of $T_b$ in both papers is $T_b = 7.6 \times 10^9 S_\mathrm{peak}(1+z) / 
(\theta_\mathrm{maj} \theta_\mathrm{min})$  where $S_\mathrm{peak}$ 
is the flux density in Jy\,beam$^{-1}$ and $\theta_\mathrm{maj}$ and
$\theta_\mathrm{min}$ are the major and minor axes of the beamwidth in 
milliarcseconds (mas).

\begin{deluxetable}{lcclcccrccc}
\renewcommand{\arraystretch}{.6} 
\tabletypesize{\scriptsize}
\tablecaption{Source Structure. \label{table:sourcestructure}}
\tablewidth{0pt}
\tablehead{
\colhead{} &
\colhead{$S_{\rm total}$} &
\colhead{Luminosity} &
\colhead{} &
\colhead{Contour} &
\colhead{$\theta_{\rm maj}$} &
\colhead{$\theta_{\rm min}$} &
\colhead{P.A.} &
\colhead{$S_{\rm peak}$} &
\colhead{$T_b$} &
\colhead{} \\
\colhead{Source} &
\colhead{(Jy)} &
\colhead{(W\,Hz$^{-1}$)} &
\colhead{Epoch} &
\colhead{(mJy)} &
\colhead{(mas)} &
\colhead{(mas)} &
\colhead{($^\circ$)} &
\colhead{(Jy\,beam$^{-1}$)} &
\colhead{(K)} &
\colhead{Structure\tablenotemark{a}}
}
\startdata
0108$+$388 & 0.43 & 6.11$\times10^{26}$ & 06/11/1999 & 1.1 & 1.13 & 0.67 & --28.0 &   0.12 & 1.93$\times10^{9}$ & SS \\
0119$+$115 & 1.37 & 1.40$\times10^{27}$ & 30/10/1998 & 2.2 & 1.09 & 0.51 &  --3.9 &   1.11  & 2.36$\times10^{10}$ & C \\
0201$+$113 & 0.65 & 3.87$\times10^{28}$ & 01/11/1998 & 0.9 & 1.15 & 0.52 &  --7.6 &   0.47 & 2.73$\times10^{10}$ & SS \\
0221$+$067 & 0.83 & 6.73$\times10^{26}$ & 06/11/1999 & 1.4 & 1.29 & 0.54 & --10.3 &   0.71 & 1.17$\times10^{10}$ & SS \\
0310$+$013 & 0.16 & 2.30$\times10^{26}$ & 01/11/1998 & 0.8 & 1.29 & 0.55 &  --6.2 &   0.11 & 1.91$\times10^{9}$ & SS \\
0405$-$385 & 1.01 & 6.13$\times10^{27}$ & 01/11/1998 & 2.4 & 1.43 & 0.44 &    1.4 &   0.79 & 2.20$\times10^{10}$ & SS \\
0420$+$022 & 1.17 & ...                  & 06/11/1999 & 2.7 & 1.37 & 0.54 & --10.6 &  0.91 & 9.31$\times10^{9}$ & SS \\
0723$-$008 & 1.01 & 4.40$\times10^{25}$ & 18/08/1997 & 1.8 & 1.03 & 0.47 &    1.1 &   0.60 & 1.06$\times10^{10}$ & SS \\
0834$-$201 & 1.99 & 6.69$\times10^{28}$ & 18/08/1997 & 2.9 & 1.09 & 0.43 &    0.6 &   1.47  & 9.01$\times10^{10}$ & SS \\
0836$+$710 & 2.24 & 4.41$\times10^{28}$ & 19/03/1998 & 4.5 & 0.80 & 0.52 & --12.3 &   1.49  & 8.58$\times10^{10}$ & SS \\
0859$+$470 & 0.63 & 5.10$\times10^{27}$ & 19/07/1999 & 1.9 & 1.14 & 0.66 &  --7.1 &   0.47 & 1.19$\times10^{10}$ & SS \\
0906$+$015 & 0.95 & 3.47$\times10^{27}$ & 18/08/1997 & 1.8 & 1.00 & 0.46 &    2.6 &   0.88 & 2.89$\times10^{10}$ & Irr \\
1032$-$199 & 1.25 & 2.54$\times10^{28}$ & 19/03/1998 & 7.2 & 1.27 & 0.50 &    0.5 &   0.91 & 3.44$\times10^{10}$ & SS \\
1124$-$186 & 2.82 & 1.09$\times10^{28}$ & 01/11/1998 & 2.6 & 1.38 & 0.50 &  --4.2 &   2.62  & 5.89$\times10^{10}$ & C  \\
1145$-$071 & 0.54 & 3.59$\times10^{27}$ & 01/11/1998 & 1.2  & 1.27 & 0.52 &  --2.2 &  0.39 & 1.06$\times10^{10}$ & SS \\
1148$-$001 & 0.87 & 1.39$\times10^{28}$ & 01/11/1998 & 0.7  & 1.17 & 0.51 &  --1.0 &  0.36 & 1.35$\times10^{10}$ & SS \\
1345$+$125 & 0.70 & 2.76$\times10^{25}$ & 01/11/1998 & 1.2 & 1.16 & 0.62 &    5.9 &   0.14 & 1.62$\times10^{9}$ & SS \\
1354$-$152 & 0.48 & 8.09$\times10^{26}$ & 19/07/1999 & 1.5 & 1.43 & 0.55 & --19.8 &   0.51 & 8.43$\times10^{9}$ & SS \\
1354$+$196 & 0.75 & 1.08$\times10^{28}$ & 18/08/1997 & 0.9 & 1.14 & 0.46 &  --1.5 &   0.43 & 1.84$\times10^{10}$ & SS \\
1458$+$718 & 1.19 & 3.33$\times10^{27}$ & 28/08/1997 & 1.2 & 0.86 & 0.51 & --13.1 &   0.62 & 2.03$\times10^{10}$ & SS \\
1502$+$106 & 1.11 & 1.49$\times10^{28}$ & 11/01/2000 & 3.0 & 1.23 & 0.53 &  --7.6 &   0.76 & 2.52$\times10^{10}$ & SS \\
1504$-$167 & 1.80 & 2.65$\times10^{27}$ & 28/08/1997 & 1.1  & 0.88 & 0.51 &  --2.9 &  0.55 & 1.55$\times10^{10}$ & SS \\
1504$+$377 & 0.73 & 1.90$\times10^{27}$ & 18/08/1997 & 2.4 & 1.16 & 0.46 &    0.0 &   1.18  & 3.13$\times10^{10}$ & SS \\
1511$-$100 & 1.34 & 1.18$\times10^{28}$ & 18/08/1997 & 4.5 & 1.20 & 0.50 &    0.1 &   1.13  & 3.61$\times10^{10}$ & SS \\
1514$-$241 & 2.23 & 1.38$\times10^{25}$ & 18/08/1997 & 6.0 & 1.16 & 0.44 &  --1.3 &   1.21  & 1.91$\times10^{10}$ & SS \\
1519$-$273 & 1.16 & 1.54$\times10^{25}$ & 18/08/1997 & 1.0 & 1.15 & 0.43 &  --2.0 &   1.00 & 1.64$\times10^{10}$ & SS \\
1538$+$149 & 0.56 & 6.56$\times10^{26}$ & 18/08/1997 & 0.8 & 0.95 & 0.52 &    0.0 &   0.42 & 1.05$\times10^{10}$ & SS \\
1555$+$001 & 0.71 & 8.90$\times10^{27}$ & 18/08/1997 & 1.1 & 1.08 & 0.50 &    1.6 &   0.56 & 2.19$\times10^{10}$ & SS \\
1622$-$253 & 2.52 & 5.19$\times10^{27}$ & 18/08/1997 & 2.3 & 1.26 & 0.46 &  --1.7 &   2.34  & 5.51$\times10^{10}$ & SS \\
1739$+$522 & 1.77 & 1.25$\times10^{28}$ & 19/03/1998 & 3.9 & 0.93 & 0.54 &  --1.7 &   1.31  & 4.70$\times10^{10}$ & C  \\
1828$+$487 & 1.96 & 3.07$\times10^{27}$ & 28/08/1997 & 2.5 & 1.15 & 0.90 &   18.8 &   1.25  & 1.56$\times10^{10}$ & SS \\
1954$+$513 & 0.94 & 5.12$\times10^{27}$ & 19/03/1998 & 2.0 & 1.02 & 0.55 &   11.6 &   0.51 & 1.54$\times10^{10}$ & SS \\
2121$+$053 & 2.50 & 3.83$\times10^{28}$ & 06/11/1999 & 4.1 & 1.34 & 0.54 &  --9.8 &   2.04  & 6.36$\times10^{10}$ & SS \\
2128$-$123 & 1.99 & 1.54$\times10^{27}$ & 18/08/1997 & 2.1 & 1.17 & 0.48 &  --0.6 &   1.04  & 2.13$\times10^{10}$ & SS \\
2128$+$048 & 0.51 & 1.75$\times10^{27}$ & 18/08/1997 & 3.4 & 1.73 & 1.19 & --10.2 &   0.17 & 1.26$\times10^{9}$  & SS \\
2155$-$152 & 1.66 & 2.43$\times10^{27}$ & 18/08/1997 & 2.2 & 1.16 & 0.47 &    0.2 &   1.10  & 2.59$\times10^{10}$ & SS \\
2227$-$088 & 0.46 & 4.29$\times10^{27}$ & 18/08/1997 & 1.4 & 1.14 & 0.48 &  --0.6 &   0.36 & 1.32$\times10^{10}$ & C  \\
2255$-$282 & 6.80 & 2.01$\times10^{28}$ & 18/08/1997 & 6.5 & 1.29 & 0.48 &    6.0 &   6.49  & 1.55$\times10^{11}$ & SS \\
2318$+$049 & 0.75 & 9.25$\times10^{26}$ & 18/08/1997 & 0.9 & 1.06 & 0.51 &  --0.7 &   0.69  & 1.57$\times10^{10}$ & SS \\
\enddata

\tablenotetext{a}{Classification of the structure, as follows: (C) compact,
(SS) single-sided, 
or (Irr) irregular.}
\end{deluxetable}

\begin{figure*}[p]
\figurenum{2}
{\psfig{figure=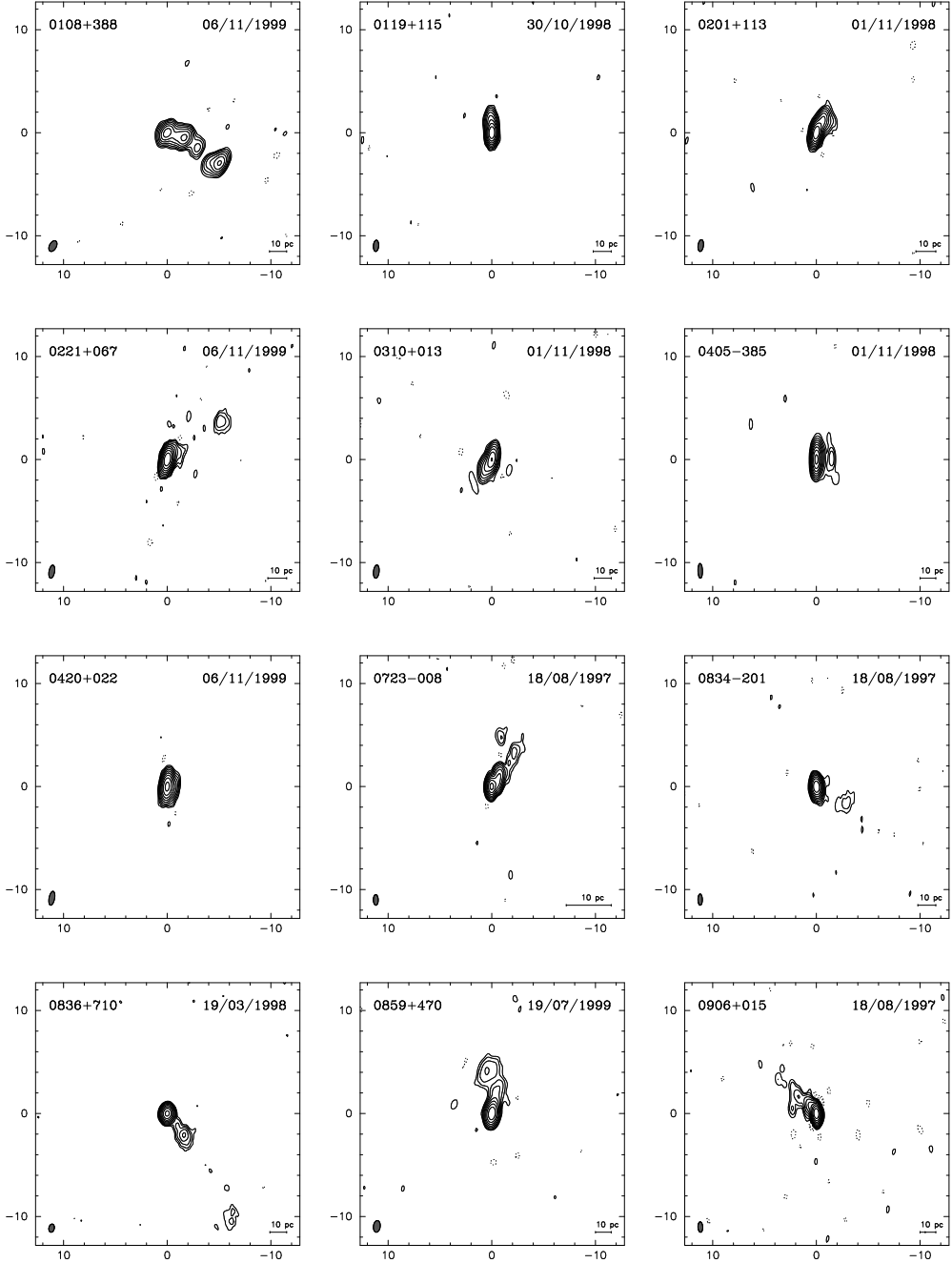,width=0.80\textwidth,angle=0}}
\caption{
Contour maps. The lowest contour is level is
generally at 3 times the root-mean-square noise and is listed in col.\ (5)
of Table~\ref{table:sourcestructure}.  The peak flux density in
each image is given in col.\ (9) of Table~\ref{table:sourcestructure}, and
the major axis, minor axis, and position angle of the restoring beam are given 
in cols.\ (6)-(8).  Most images are centered on the brigthest component,
but for a few of the larger asymmetric sourcs, we have shifted the center
to fit the image in the 25.6\,mas box.  Note that while most of the displayed
boxes are 25.6\,mas on a side, a few of the larger sources are shown in
a box twice this size.  Each panel also shows a bar representing a linear
scale of 10\,pc.
}
\label{fig:maps}
\end{figure*}

\begin{figure*}[p]
\figurenum{2}
{\psfig{figure=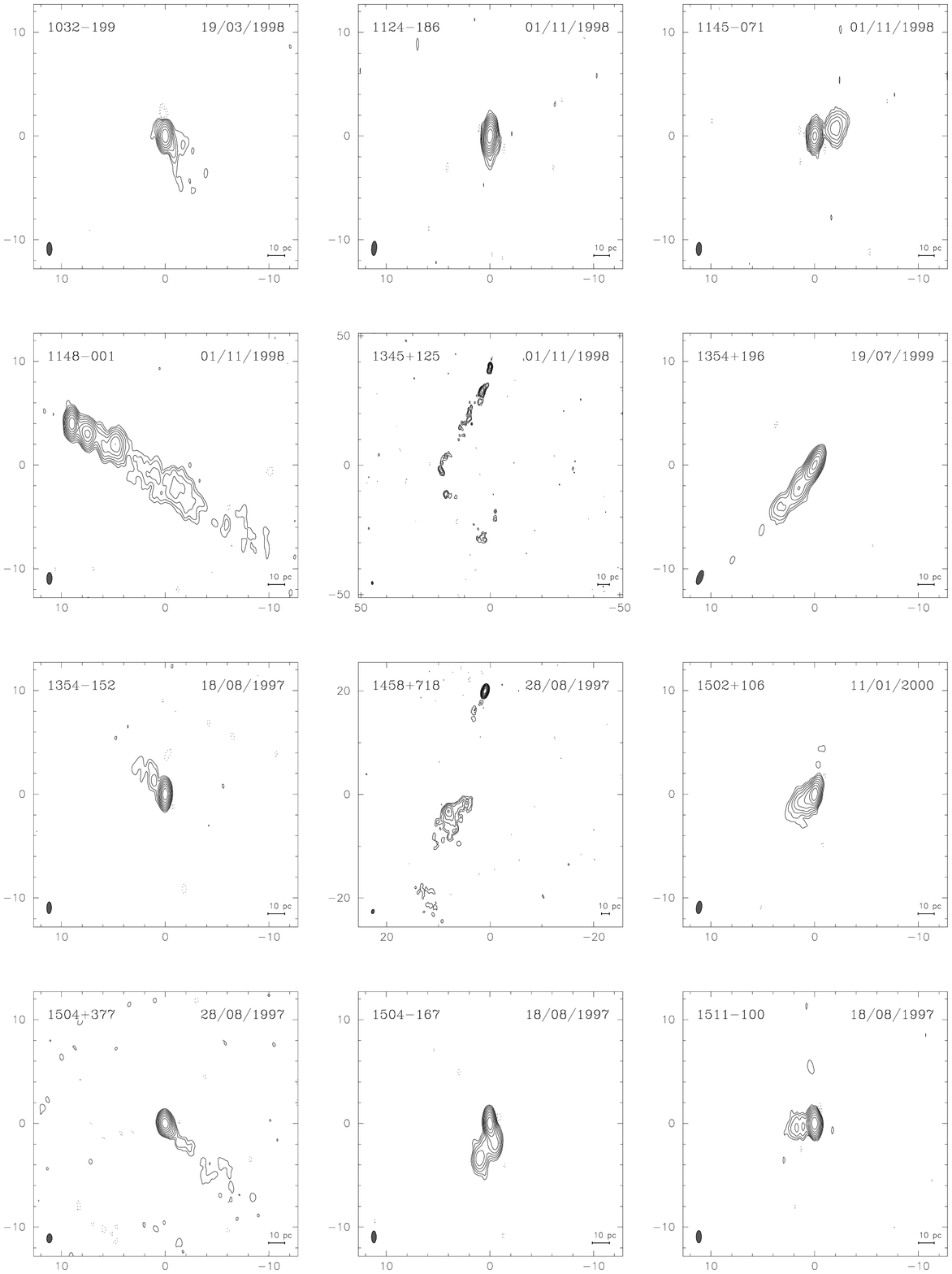,width=0.80\textwidth,angle=0}}
\caption{{\em Continued}}
\label{fig:fig2}
\end{figure*}

\begin{figure*}[p]
\figurenum{2}
{\psfig{figure=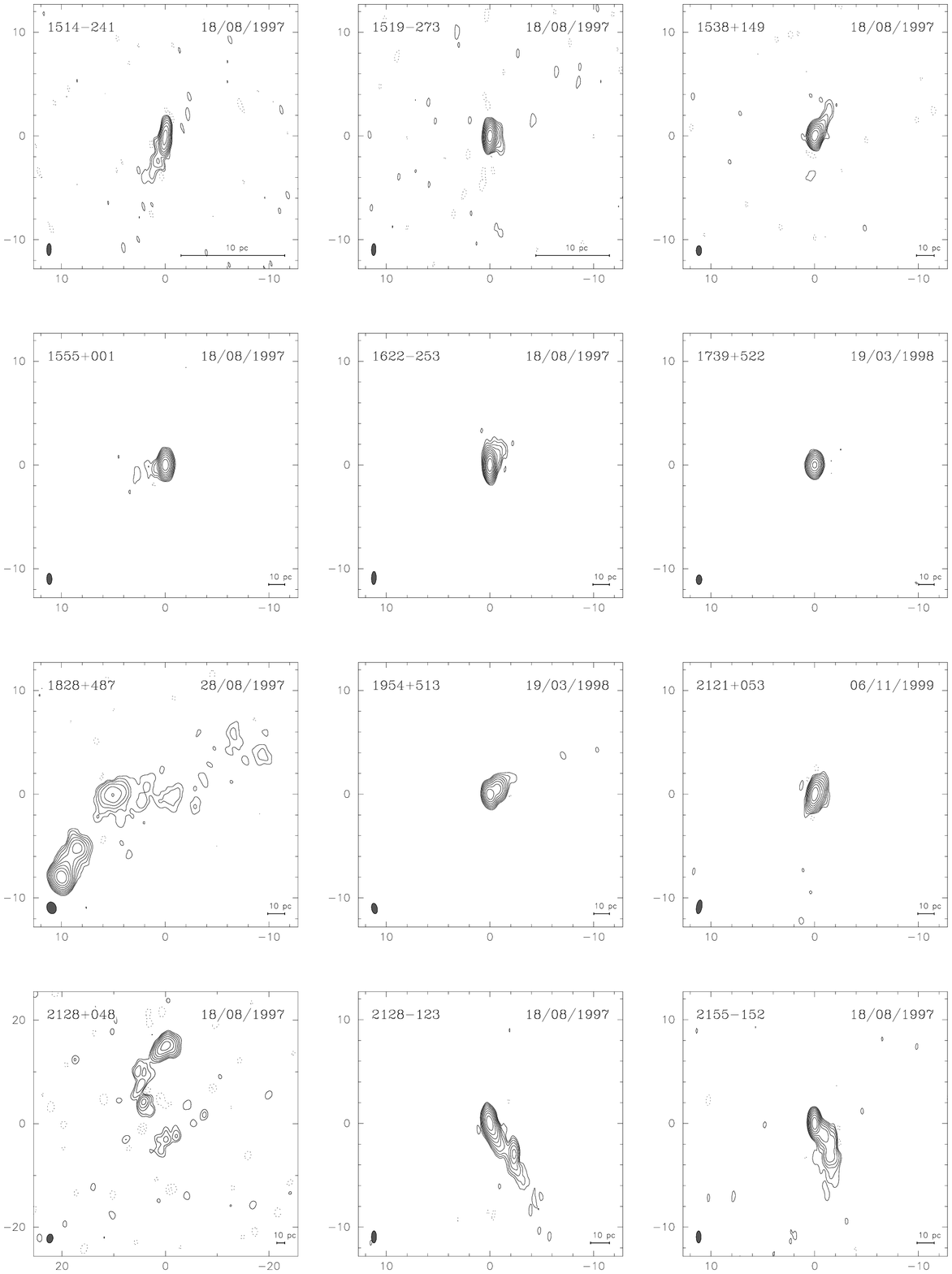,width=0.80\textwidth,angle=0}}
\caption{{\em Continued}}
\label{fig:fig3}
\end{figure*}

\begin{figure*}[p]
\figurenum{2}
{\psfig{figure=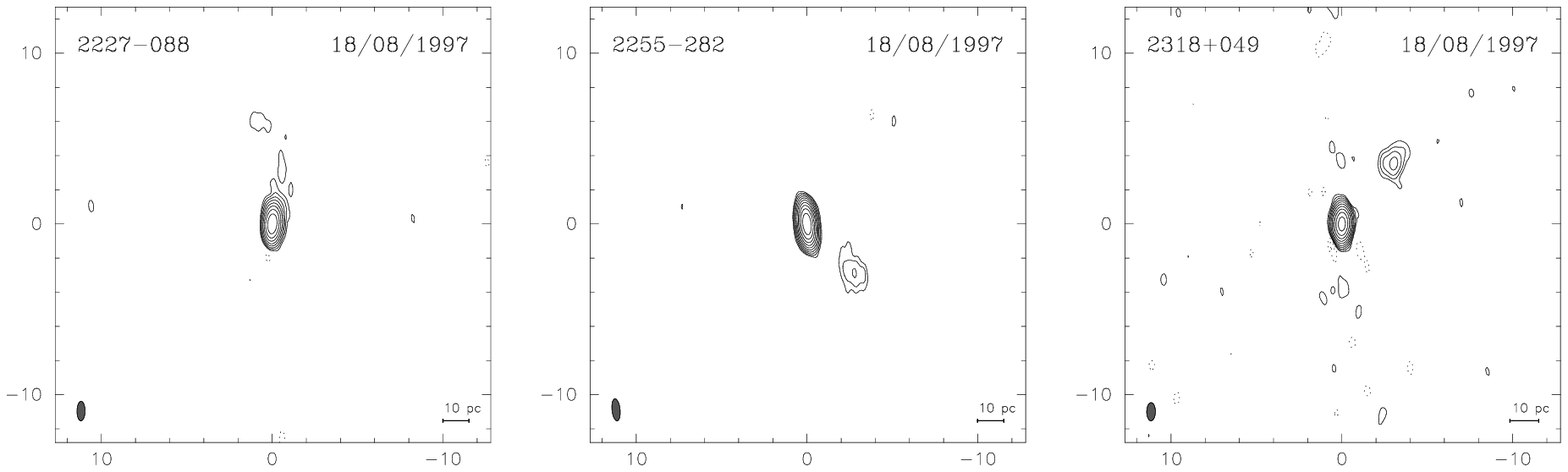,width=0.80\textwidth,angle=0}}
\caption{{\em Continued}}
\label{fig:fig4}
\end{figure*}

\subsection{Morphology \label{subsec:morphology}}

The last column in Table~\ref{table:sourcestructure} gives the morphological
classification, as follows. C (compact) refers to sources that are unresolved,
or barely resolved. SS refers to sources that appear to be single-sided; i.e.\
the presumed core, identified by compactness, is at one end of the brightness
distribution. 
Irr (irregular) refers to those sources where the
structure appears two-dimensional.  This classification scheme is the same
as that used in Paper I, where it is discussed in more detail.

Table~\ref{table:sourcedistribution} summarizes the classification for 169 of
the combined set of 171 sources from this paper and from Paper I.  Two sources
in Paper I could not be classified: 0218+357 which is a lensed object
(\citealt{Patnaik93}), and 1328+254 which is nearly fully resolved.
Objects near zero declination may have low-level artifacts located to the
north or south of the real brightness distribution, and in a few cases the
classification is ambiguous.  See Section \ref{sec:notesindiv}.

\begin{deluxetable}{lcc}
\tabletypesize{\footnotesize}
\tablecaption{Distribution of morphology of the
complete survey\label{table:sourcedistribution}}
\tablewidth{0pt}
\tablehead{
\colhead{Structure\tablenotemark{a} } & 
\colhead{Number} & 
\colhead{\%} 
}
\startdata
SS  &  142 &  83 \\
C   &   13 &   8 \\
DS  &    8 &   5 \\
Irr &    6 &   4 \\
\enddata

\tablenotetext{a}{Classification of the structure, as follows: (C) compact,
(SS) single-sided, 
(DS) double-sided, 
or (Irr) irregular.}
\end{deluxetable}

Most of the sources appear single-sided, which is
probably the result of Doppler
favoritism: the approaching jet is relativistically-boosted and the
receding jet is weakened and thus typically not detectable. As
expected, all sources from the complete sample which show superluminal internal motions are in this
category. Sources which show sub-luminal motions may have either a
single-sided or a double-sided appearance. The internal motions will be 
discussed in detail by Kellermann et al.\ (in preparation).

\subsection{Brightness Temperature \label{subsec:t_b}}

In many cases the compact
components are unresolved by the 15\,GHz VLBA synthesized
beam. Most of them may be associated with the central structure, or core.
The brightness temperature ($T_b$) in column (10) of
Table~\ref{table:sourcestructure} is calculated using the restoring beam
area, and as such is a lower limit. The existence of much smaller
entities with apparently very high values of $T_b$ is evidenced by intra-day
variability (IDV) studies (\citealt{Kraus99,Quirrenbach00,KedzioraChudczer01})

\begin{figure}[htbp]
\figurenum{3}
{\psfig{figure=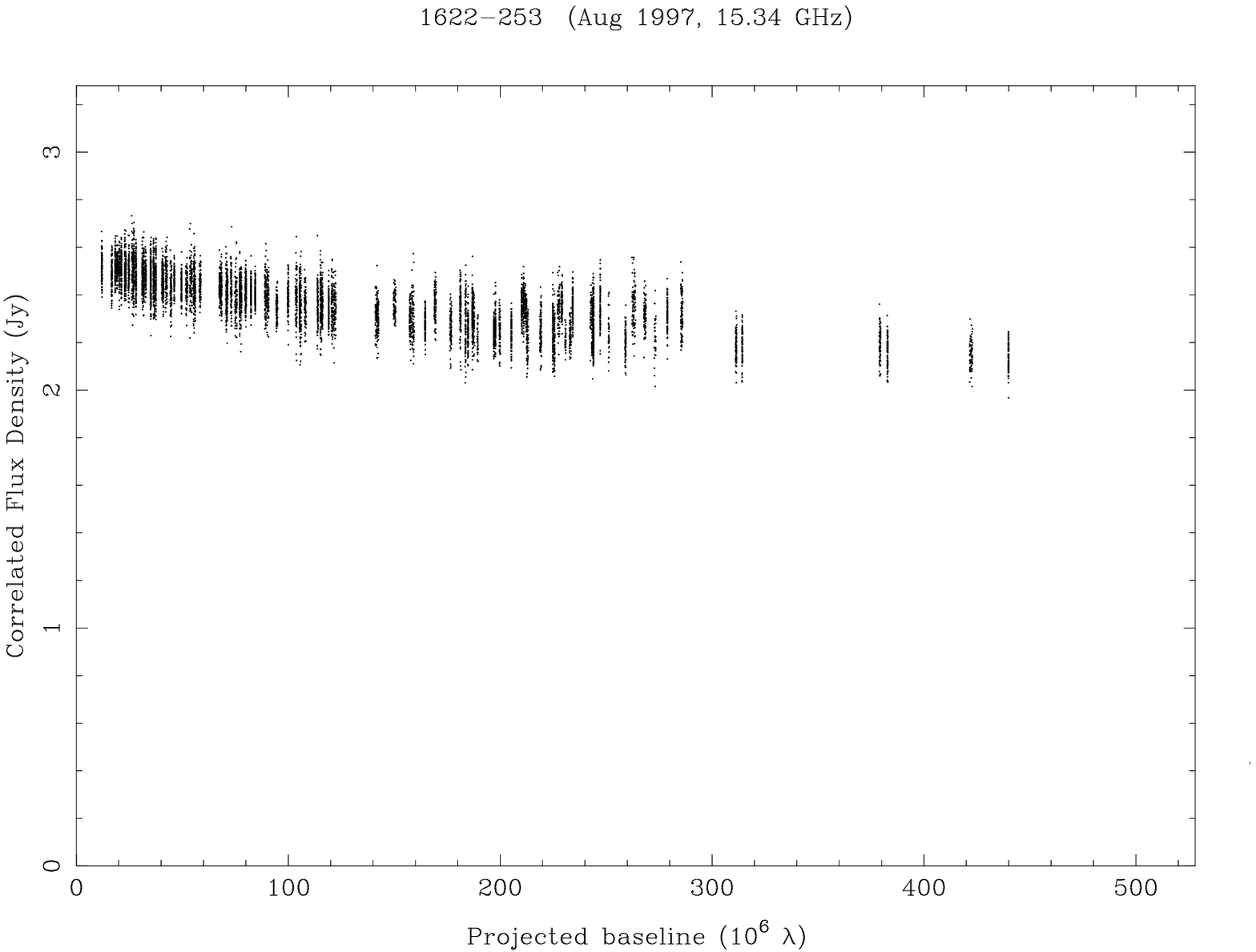,width=0.50\textwidth,angle=0}}
\caption{Visibility amplitude versus $(u,v)$ radius for the radio
source 1622--253.  
\label{fig:1622-253_radplot}
}
\end{figure}

A better estimate of component sizes which are smaller than
the restoring beam size can be obtained by 
fitting elliptical Gaussian models to
the visibility data. An example is given in Figure~\ref{fig:1622-253_radplot},
where the visibility for the source 1622--253 is plotted as a function of
$(u^2+v^2)^{1/2}$. In this case the visibility is reduced by only 
$\sim$15\%\ at 400 million wavelengths; this source can be modeled by 
two elliptical Gaussian components: one with 2.41\,Jy,
major axis 0.14\,mas, and axial ratio of 0.59
(P.A.\,21$^\circ$), giving $T_{b,{\rm vlba}}\sim 3\times10^{12}$\,Kelvin;
the other component is circular and has a flux density of
0.08\,Jy and a diameter of
0.4\,mas at a distance of 0.79\,mas 
(P.A.\ $-55^\circ$) from the compact component.

VLBA data are generally of high quality: the noise is low
(typically less than 0.5\,mJy\,beam$^{-1}$), the $(u,v)$-coverage is
good, and the relative fringe visibility, after self-calibration,
accurate to a few percent.  
Therefore we feel justified in
calculating models in this way, and using the small sizes to derive
brightness temperatures.
Table~\ref{table:t_b_flat} shows $T_{b,{\rm vlba}}$
estimated in this way, for sources that show 
nearly flat visibility plots.  These
temperatures are calculated in the AGN reference frame. They still must be
considered lower limits, since a source like that in
Figure~\ref{fig:1622-253_radplot} which drops slowly to 85\% at the limit of our
resolution can have one or more very small, possibly separated components.



\subsubsection{Comparison of VLBA and Variability Brightness Temperatures
\label{subsec:VLBA-var}}

Flux density variations provide an alternative approach to estimating
$T_b$, independent of interferometry measurements. This uses the
causality argument, that the time scale $\tau$ of a variation cannot be
less than the light-crossing time of the varying emission region.
Taking the two time scales to be equal, $\tau=r/c$, gives the maximal
allowed source radius $r$, and hence the minimal allowed brightness
temperature, $T_{b,\mathrm{var}}$, from the 
solid angle $\Omega=\pi r^2/D$, where $D$ is the
angular size distance to the source. The variability time
scale is found from the logarithmic
derivative
$\tau=\mathrm{d}t/\mathrm{d}(\ln S(t))$ (\citealt{Burbidge74}).
In this section we compare our measured brightness temperatures with those
derived by \cite{Lahteenmaki99b} from flux density
variations at 22 and 37\,GHz.

Three coordinate frames are involved with these brightness
temperatures, and to avoid confusion we use a superscript * for the
frame of the relativistically moving plasma; we use no superscript for
the frame of the AGN, and we will not define any quantities in the
local coordinate frame. The transformations of 
$T_{b,\mathrm{vlba}}$ and $T_{b,\mathrm{var}}$ among
these coordinate systems have different powers of $(1+z)$ and $\delta$ (the
Doppler factor of the moving plasma). With several
assumptions, this allows us to calculate the intrinsic 
brightness temperature 
$T^*_{b,\mathrm{int}}$. We have two values of 
$T^*$:  $T^*_{b,\mathrm{vlba}}=\delta^{-1}\times T_{b,\mathrm{vlba}}$, 
and $T^*_{b,\mathrm{var}} = \delta^{-3}\times T_{b,\mathrm{var}}$. 
Strictly speaking,
$T^*_{b,\mathrm{vlba}}$ and $T^*_{b,\mathrm{var}}$ 
refer to different plasmas, because the measurements
are at different wavelengths, and at different epochs. 

In simple models of relativistic conical jets (e.g., \citealt{Koenigl81})
the radio radiation from the core comes mainly from the region of the
jet where the optical depth is unity. This varies with wavelength
because the density and magnetic field decrease with distance from the
apex. But the velocity changes slowly if at all, and hence the Doppler
factor can be assumed to be the same for all the frequencies
we are using (15, 22, and 37\,GHz), which are separated by
a factor of only 2.5.
The kinetic temperature, which
controls the brightness temperature when the optical depth is unity,
probably does change slowly along the jet; the biggest effect might be
the sideways expansion of the plasma as it moves out. This effect
likely is small over a frequency range of 2.5; and indeed the kinetic
temperature has been taken as constant in various studies of jets 
(e.g.\ \citealt{Zensus95}).

The fact that the data for a given source come from different epochs is of
concern, but we assume that this is not important. The VLBA measurement
refers to the core which seems to be a permanent feature of these
sources, and not to features in the jet which are evanescent.  And, the
variability time scale appears to change little between outbursts in
any one source (\citealt{Valtaoja99}).

Thus, we assume that estimates of kinetic temperature derived from the
VLBA and from the variability observations should be approximately the same. 
This is valid only for observations of the flat-spectrum core;
it is not valid for the outer steep-spectrum components. Assuming
$T^*_{b,\mathrm{vlba}} = T^*_{b,\mathrm{var}}$, 
we derive formulas for estimating the intrinsic
temperature $T^*_{b,\mathrm{int}}=T^*_b$ and the Doppler factor:
\begin{equation}   
       T^*_{b,\mathrm{int}} = \sqrt{T_{b,\mathrm{vlba}}^3/T_{b,\mathrm{var}}},  
                                                      \label{eq:tbint} 
\end{equation}
\begin{equation}
             \delta = \sqrt{T_{b,\mathrm{var}}/T_{b,\mathrm{vlba}}}.        \label{eq:D} 
\end{equation}
Equation (\ref{eq:tbint}) is the same as Eq.\ (8) in \citet{Lahteenmaki99a}.

Table \ref{table:t_b_flat} shows $T_{b,\mathrm{var}}$ from
(\citealt{Lahteenmaki99b}; Table~1) with 
values converted to
$H_0$=65\,km\,s$^{-1}$\,Mpc$^{-1}$, $q_0$=0.5, for those sources which 
have both a VLBA and a
variability brightness temperature. Columns 4 and 5 of Table
\ref{table:t_b_flat} contain $T^*_{b,\mathrm{int}}$ and $\delta$,
respectively, calculated from Eqs.\ (\ref{eq:tbint}) and (\ref{eq:D}).
We note first that in every case $T_{b,\mathrm{var}} >
T_{b,\mathrm{vlba}}$. This is required for 
$\delta\geq 1$, so in this sense, the data are self-consistent.  Values
of $\delta$ range from barely relativistic for 0007+016 to
$\delta\sim50$ for 0235+164 and 0804+499. 0235+164 is a known gamma-ray
source, and various theories for gamma rays require high values of
$\delta$ (\citealt{Stecker72,Mattox93}).
Only four sources in Table~\ref{table:t_b_flat} have EGRET
measurements.
No general conclusions can be drawn, but the sets of values are
consistent with the idea that there is a high Doppler factor in
gamma-ray sources.

It perhaps is surprising that the values obtained on
the whole correspond well with expectations, since both
$T_{b,\mathrm{var}}$ and $T_{b,\mathrm{vlba}}$ are lower limits,
creating an extra margin of uncertainty in the derived values of
$T^*_{b,\mathrm{int}}$ and $\delta$.

\citet{Lahteenmaki99a} make a similar analysis and show a plot of
$T_{b,\mathrm{vlba}}$ against $T_{b,\mathrm{var}}$, with diagonal lines
indicating values of constant $T^*_{b,\mathrm{int}}$. Our results (Table
\ref{table:t_b_flat}) are in agreement with theirs: the points lie
between $T^*_{b,\mathrm{int}} = 10^{10}$ and $10^{12}$.

For many years a value of $10^{12}$\,K has been taken as a limit to
$T^*_{b,\mathrm{int}}$ for a synchrotron radio source in equilibrium,
because higher temperatures lead quickly to the ``inverse-Compton
catastrophe" which quenches the plasma and reduces its temperature to
$10^{12}$ (\citealt{Kellermann69}).  More recently, \citet{Readhead94}
has suggested that the equipartition temperature $T_\mathrm{eq} \sim
10^{10}$ or $10^{11}$ is more likely to be appropriate in most
objects.  Table \ref{table:t_b_flat} shows that
$T^*_{b,\mathrm{int}}=10^{12}$ is approached but not exceeded.  The
distribution of $T^*_{b,\mathrm{int}}$ has a median value $T^*_{b,\mathrm{int}}
\approx 10^{11}$, suggesting that they are close to equipartition.

\begin{deluxetable}{lrrrrc}
\tabletypesize{\footnotesize}
\tablecaption{Brightness temperatures and Doppler factors of the most
compact sources \label{table:t_b_flat}}
\tablewidth{0pt}
\tablehead{
\colhead{Source}  &
\colhead{$\log T_{b,\mathrm{vlba}}$} &  
\colhead{$\log T_{b,\mathrm{var}}$} &   
\colhead{$\log T^*_{b,\mathrm{int}}$} &
\colhead{$\delta$} &
\colhead{EGRET?} 
}
\startdata
0007$+$016 &  11.26  & 12.21  & 10.79 &   3.0 &        \\
0048$-$097 &  11.83  &        &       &       &        \\
0235$+$164 &  11.28  & 14.71  &  9.57 &  51.9 &    Y   \\
0333$+$321 &  11.68  & 13.50  & 10.77 &   8.1 &    N   \\
0607$-$157 &  12.18  &        &       &       &        \\
0642$+$449 &  12.04  & 14.75  & 10.69 &  22.6 &        \\
0804$+$499 &  11.96  & 15.32  & 10.28 &  47.9 &        \\
0808$+$019 &  11.86  &        &       &       &        \\
0906$+$015 &  12.48  &        &       &       &      \\
1253$-$055 &  12.76  & 14.74  & 11.77 &   9.8 &    Y   \\
1308$+$326 &  11.73  & 14.24  & 10.48 &  18.0 &        \\
1354$-$152 &  11.73  &        &       &       &       \\
1622$-$253 &  12.18  &        &       &       &    Y   \\
1638$+$398 &  11.30  &        &       &       &       \\
1741$-$038 &  12.08  & 13.92  & 11.16 &   8.3 &       \\
1749$+$096 &  12.74  & 14.67  & 11.78 &   9.2 &       \\
1758$+$388 &  11.60  &        &       &       &       \\
1921$-$293 &  11.89  &        &       &       &       \\
2255$-$282 &  12.30  &        &       &       &    Y  \\
\enddata
\end{deluxetable}

Our ground based VLBA observations are limited to an angular
resolution of about 0.15\,mas.  For radio sources of a few Janskys,
this corresponds to a measured brightness temperature
of about $10^{12}$\,K and our determination of
$T_{b,\mathrm{vlba}}$ is only a lower limit.
The variability method is limited by time
resolution; in the data we have quoted this is about 1 month,
corresponding to a maximum measurable temperature of about $10^{15}$\,K
for $z$=1.  Faster flares, corresponding to higher temperatures, are
not measurable. These limits are independent, and thus it is
particularly interesting that the values in Table 4 are rather close to
the values expected on the basis of synchrotron theory, and the
observations of superluminal motion and X-ray flux. This agreement
might mean that the true temperatures are not much above the limits.
On the other hand, it may merely reflect a coincidence: 0.1\,mas at $z$=1
corresponds to 2 light years, which gives an observed light crossing
time of about 1 month if the Doppler factor is 20. In this view the
temperatures and Doppler factors in Table 4 are largely given by the
observational limits on time and spatial resolution.  It is 
important to test these ideas by improving the limits.

\citet{Lahteenmaki99b} actually turn 
this argument around, by assuming a
particular value for $T^*_{b,\mathrm{int}}$ to calculate $\delta$.  Our
procedure is more powerful for the objects in common. But while we
believe we have the requisite high-quality high-resolution
interferometry data, we still rely on the fundamental assumptions
that the actual brightness temperatures do not differ grossly from the
limits obtained, and that $T^*_{b,\mathrm{var}} \simeq T^*_{b,\mathrm{vlba}}$.


\section{Notes on individual sources \label{sec:notesindiv}}

In this section we describe the morphological structure. 
``RRFID" indicates the observations at 2.3 and
8.4\,GHz from the USNO Radio-Optical Reference Frame 
image database (\citealt{Fey96, Fey97, Fey00}).  ``VPLS" indicates
that the source was observed as part of the 6\,cm VSOP Pre-Launch
Survey (\citealt{Fomalont00}).

\paragraph{0108+388}
This source
is a Compact Symmetric Object. Proper motions
permit an estimated kinematic age of 370\,yr
(\citealt{Owsianik98}). 
The radio flux density is weakly
polarized (0.30$\pm$0.08\% at 4.8\,GHz) without significant variations
(\citealt{Aller92}).

\paragraph{0119+115}
This source 
is unresolved by 5\,GHz polarization observations
(\citealt{Gabuzda99}).  
We observed the source at one epoch and it is slightly elongated to the 
north.  No flux is detected below 2\,mJy\,beam$^{-1}$ beyond 2\,mas from
the brightness peak.

\paragraph{0201+113}
This system has been extensively studied in the optical
due to the peculiar spectroscopic characteristics of
a Damped Lyman\,$\alpha$ System towards the host galaxy
(\citealt{Ellison01,Kanekar97,Oya98}),
H{\sc i} was detected in the radio (\citealt{DeBruyn96}).
RRFID shows structure westward at
8.4\,GHz and diffuse structure to the southwest at 2.3\,GHz.
In our images there is an elongated structure to the northwest, not
extending beyond 2\,mas.

\paragraph{0221+067}
This is a flat spectrum, highly polarized QSO.
Our image shows a component 6.3\,mas to the northwest of the core.
The structure
at lower frequencies from RRFID is also elongated in this direction.

\paragraph{0310+013}
This compact QSO shows a halo to the west at 4$^{\prime\prime}$ 
in VLA images 
published by \citet{Price93}.
In our 15\,GHz image the source has a compact 
structure elongated to the southeast.

\paragraph{0405--385}
This source is a well known IDV source
(\citealt{KedzioraChudczer96,KedzioraChudczer97,KedzioraChudczer98}), 
with periods of 50\% variability
in one hour.
Circular polarization of --0.101\% was reported by
\citet{Rayner00}.
Our 1998 image shows a component to the west at 1.5\,mas, compatible
with the RRFID images.

\paragraph{0723--008}
This BL Lac type object 
has a galactic latitude of $+7^\circ$.
The RRFID  images show 
a jet extending to the northwest at 8.4\,GHz up to 7\,mas
and turns to the north at distances of 40\,mas (2.3\,GHz image).  
This structure
is also seen on the 1.6\,GHz images of \citet{Bondi96}.
Our image from August 1997 shows a clear distinct component at 
2\,mas to the northwest and more diffuse emission extending to 6\,mas from 
the core.

\paragraph{0834--201}
This source is a flat-spectrum
blazar (\citealt{Hewitt89}).
The first VLBI observations of this
radio source were carried out 
at 1.6\,GHz by \citet{Kellermann71}.
The VLBA calibrator survey 
images at 2.3\,GHz shows a compact structure
extending to the west (\citealt{Beasley96}). 
Our images show a compact source with a very faint component
4\,mas away in position angle (P.A.) $-100^\circ$.

\paragraph{0836+710}
This radio source has a highly-polarized
secondary component at a distance
of 1.3$^{\prime\prime}$ from the core along P.A.\ 200$^\circ$
(\citealt{Perley82}).  
MERLIN-VLBI data show a jet structure
extending up to 150\,mas in the direction to the outer arcsecond
lobe (\citealt{Hummel92}).  
The source has been monitored on pc-scales and shows a complex 
one-sided core-jet structure along P.A.\ 215$^\circ$.  
(\citealt{Krichbaum90,Otterbein98,Lobanov98,Hutchison01,Ros01,Lister01}).
The jet follows a slightly curved path where components travel
outwards at speeds of $\sim0.3$\,mas/yr ($\sim10c$).
Our images reveal a bright component 2\,mas away from the core, and jet
extended out to 12\,mas.
The brightness evolution reported by \citet{Peng00} shows a slight rise by
the mid 90's reaching a peak around 1997 and decaying afterwards.
Optical flares and X-ray flares have also been observed 
(\citealt{VonLinde93,Malizia00}).

\paragraph{0859+470}
This source has been observed with VLBI since the 1980's
(e.g. \citealt{Zensus84,Lawrence85,Pearson88,Lister01}, etc.).
The RRFID images
show an elongated, diffuse
structure extending north up to 100\,mas at 2.3\,GHz.
A 43\,GHz VLBA image by \citet{Lister01b} shows a curved jet
structure one milliarcsecond from the core. 
Our images from July 1999 shows two clear components, at 2 and 4\,mas
from the core, in P.A.\,10$^\circ$ and $-10^\circ$, respectively.

\paragraph{0906+015}
The kpc-scale image from this flat-spectrum QSO published
by \citet{Murphy93} show a bright component
12$^{\prime\prime}$ east of the core.
The RRFID images show a jet towards the northeast, reaching
30\,mas at 2.3\,GHz.
Our images show jet structure at the same direction, extending out to
3--4\,mas.

\paragraph{1032--199}
To our knowledge, no VLBI images from this source have been published before.
Our image shows a bright core and some diffuse structure towards the southwest.

\paragraph{1124--186}
The images and visibilities from the RRFID
and from our data show a compact, unresolved source.

\paragraph{1145--071}
\citet{Djorgovski87} 
and \citet{Kochanek99} 
reported the source as a probable
binary quasar, with two optical features at the same
redshift and separated by $4\rlap{.}^{\prime\prime}2$ in projection.
Our image shows a secondary component at the same distance (2.1\,mas)
and position (P.A.\ $-63^\circ$) as the RRFID image.

\paragraph{1148--001}
\citet{Kellermann71} reported a compact structure
of this source at pc-scales, and Inter Planetary Scintillation (IPS)
observations suggested
three compact components at kpc-scales (\citealt{Venugopal85}).
The RRFID images show a jet to the southwest turning to the south
beyond 20\,mas of the core, extending up to more than 40\,mas at 2.3\,GHz,
and with 
a bright component at 3.2\,mas, P.A.\,$-119^\circ$,
and a more extended one at 10.9\,mas, P.A.\,$-123^\circ$ at 8.5\,GHz.
Our image from November 1998 reports a conspicuous jet extending to the
southwest over 20\,mas, with one component located 2 and other at 5\,mas from the core.

\paragraph{1345+125}
The extremely reddened ($U$--$B$=0.73, $B$--$I$=2.54) 
Seyfert 2 host galaxy 
contains two nuclei separated by $2^{\prime\prime}$
(P.A.\,$-75^\circ$)
suggesting an ongoing merger (\citealt{Gilmore86}).  
\citet{Axon00} claim that the radio source is associated with the
western component of the optical double nucleus from HST observations.
This object is similar to Arp\,220, the prototype mega-maser galaxy.  
The Arecibo flux density monitoring at 318 and 430\,MHz 
\citep{Salgado99} shows no variations over a 14-year period.
The structure of 1345+125 is discussed in more detail by Lister et al.\ 
(in preparation).

\paragraph{1354+196}
This source was observed at 22\,GHz by \citet{Moellenbrock96}
in the VSOP pre-launch survey.
Our images show 
a core-jet structure to the southeast, with components
at 2 and 5\,mas from the core, consistent with RRFID results.

\paragraph{1354--152}
Our images show compact structure elongated to the NE,
with diffuse emission out to $\sim$3\,mas.

\paragraph{1458+718}
This is a prominent compact
steep-spectrum radio source (\citealt{VanBreugel84,Fanti95}).  
\citet{Aaron97,Aaron98} studied the polarization 
and multi-frequency properties.
Our image from August 1997 reveals a compact core and a more
extended component 20\,mas to the south, with some extended,
diffuse emission.  Lower-frequency images show
emission connecting those two bright features and extending 
to 50\,mas from the core before turning to the east (\citealt{Aaron98}).

\paragraph{1502+106}
\citet{Hardcastle97} presented 1.5 and 8.4\,GHz observations of
this FRI radio galaxy. Their observations show
a broad two-sided jet emanating from a
bright, compact core, at  P.A.$\sim$$30^\circ$.  
\citet{Martel99} show HST images with a sub-kpc, elongated
and dusty structure centered on its nucleus
at P.A.\ $163^\circ$.  The radio jet
is orthogonal to the major axis of the dusty structure.
The RRFID images display a diffuse pc-structure 
5--10\,mas to the east of the compact core.
Our images show a 3--4\,mas structure extending to the
southeast.

\paragraph{1504+377}
\citet{Carilli97} reported H{\sc i} absorption
lines toward this inclined disk galaxy.
\citet{Wiklind96} found
seven different molecular lines toward this object.
1.6\,GHz VLBI images by \citet{Polatidis95} show an extended jet 
100\,mas to the southwest.
The RRFID images show structure up to
70\,mas at 2.3\,GHz, and 15\,mas at 8.4\,GHz.

\paragraph{1504--167}
5\,GHz VLBI images from this highly polarized quasar (\citealt{Shen97}) 
show a compact structure with
three resolved components within 1--2\,mas from the core.
The RRFID data show a compact structure elongated
southwards at 8.4\,GHz.
Our images show a non-linear three-component morphology, with the core,
one component at $\sim$2\,mas, P.A.\ -160$^\circ$, and 
one at 4\,mas, P.A.\ 160$^\circ$ from the core.  


\paragraph{1514--241}
This nearby N-type galaxy
displays high variability in the optical, reaching rates of
0.06$\pm$0.01\,mag\,hr$^{-1}$ (\citealt{Carini91,Bond71}).
The HST images published by \citet{Scarpa00} show a bright point surrounded
by a large, round elliptical host.
It has a pc-scale straight radio jet extending at least
20$^{\prime\prime}$ to the east of the core at P.A.\,88$^\circ$
(\citealt{Morganti93}).
\citet{Lister98} published 43\,GHz polarimetric images of this
object, showing a jet pointing towards P.A.\,$-171^\circ$, 
within 1\,mas of the core, then turning to P.A.\,157$^\circ$.
The RRFID images show a jet towards the southeast.  Our image
shows structure towards the southeast, out to 5\,mas.

\paragraph{1519--273}
This BL\,Lac object 
has a featureless spectrum, and a high degree (5--12\%) of
variable optical linear polarization (\citealt{Impey88}).
It displays IDV at radio frequencies,
and also in circular polarization (\citealt{Macquart00}).
It is unresolved at kpc-scales (\citealt{Cassaro99}).
Our image shows a very compact structure with slight extension
to the west.

\paragraph{1538+149}
This is an optically variable BL\,Lac object.
\citet{Ulvestad83} report an unresolved structure within 4$^{\prime\prime}$
at kpc-scales, with hints of extension northwards.
The RRFID images show emission towards the 
northwest up to 60\,mas at 2.3\,GHz
and up to 10\,mas at 8.4\,GHz.
Our image shows only the inner part of this elongated structure.

\paragraph{1555+001}
Our images show a very compact structure with some emission to the east,
extending up to 5--6\,mas at the 1\,mJy\,beam$^{-1}$ level.

\paragraph{1622--253}
\citet{Morganti93} report an unresolved radio structure at kpc
scales.  The RRFID images show diffuse
emission to the north and northwest, without identifiable components from
one epoch to another.  Our images show a compact structure 
extending up to 3\,mas to the northwest at the 1.5\,mJy\,beam$^{-1}$ level.  

\paragraph{1739+522}
A MERLIN image of this highly polarized QSO shows a secondary component 
3$^{\prime\prime}$ away from the core (\citealt{Reid95}).
A high-resolution 43 GHz image of the core
region by \citet{Lister01b} shows a jet that starts out in an
easterly direction and curves over 90$^\circ$ to the north. 
Our image shows only an unresolved structure.

\paragraph{1828+487}
This is a powerful FRII CSS radio source
surrounded by a halo of 14$^{\prime\prime}\times$9$^{\prime\prime}$ in
size
(\citealt{VanBreugel92}).  
The superluminal motions in this source are described
by \citet{Polatidis98}.
\citet{Taylor98} reported high rotation measures at parsec scales.
The source has a prominent jet extending towards the northwest, which is
also visible in our 
image. Components can be identified at distances
of $\sim$4 and $\sim$10\,mas 
from the core in P.A.\ $\sim$30$^\circ$.  
Similar structures are visible in the
5\,GHz space-VLBI image of \citet{Lister01}.

\paragraph{1954+513}
The 5\,GHz VLA images from \citet{Kollgaard90} shows a north-south double-sided
jet, with two components at 5 and 10$^{\prime\prime}$ north of the core,
and one at 7$^{\prime\prime}$ to the south.
Our image shows a well-defined jet that extends for approximately 
10\,mas to the northwest at P.A.\ $70^\circ$, which agrees with a
recent 5 GHz space-VLBI image by \citet{Lister01}.
This jet extends more than 80\,mas at 2.3\,GHz in the RRFID
images. 

\paragraph{2121+053}
Over the past decades, 
data from the UMRAO show a slow variation in flux density 
of this source at 15\,GHz 
(e.g.\ \citealt{Aller92}).
The kpc-scale images from \citet{Murphy93} show an unresolved source with
hints of emission to the east.
Our images show a very compact structure, elongated to the west.

\paragraph{2128--123}
This is a radio-variable, flat-spectrum QSO.
Our images show the same highly curved 
jet structure
as in the RRFID images
with a prominent component located 3.7\,mas from the core.

\paragraph{2128+048}
This very red and weak 
galaxy has a radio spectrum which is peaked near 600\,MHz.
The 8.4\,GHz pc-scale images from \citet{Dallacasa98} report 
a triple component structure with a possible tail 
at 2.3\,GHz, compatible with the 
images from \citet{Stanghellini97}.  
Our image shows a core-jet structure with an elongated core to the southeast and
with
further components at 10\,mas (P.A.\ $\sim$160$^\circ$) and 20\,mas
(P.A.\, 180$^\circ$) from the core.

\paragraph{2155--152}
On kpc-scales this source displays a triple structure
extending 6$^{\prime\prime}$ in the north-south direction (\citealt{Perley82}).
The 5\,GHz images published by \citet{Shen98} show a
compact core-jet structure
aligned with the kpc-scale image.  Our image shows a jet towards the
south with a bright jet and three components up to 7\,mas away.

\paragraph{2227--088}
Our images show a very compact core with some faint emission 
extending towards the north,
up to 6\,mas at 1\,mJy\,beam$^{-1}$, compatible with the RRFID images.

\paragraph{2255--282}
This is an optically variable radio source.  
A gamma-ray flare was detected from EGRET
in 1997 (\citealt{Macomb99}).  It has been studied at different 
radio frequencies
and scales by \citet{Tornikoski99} before and after the flare.  They
present a 5\,GHz VLBA image with a jet pointing 
towards the southwest (also seen in the RRFID images), but their 
higher frequency observations show only an unresolved source.
Our images show a compact core and a component at P.A.\,$\sim$$-120^\circ$.

\paragraph{2318+049}
VLA images (\citealt{Hutchings98}) show an unresolved structure
slightly elongated along P.A.\,$-40^\circ$.
Our images also show a core-jet structure oriented  in
the same direction with one distinct component located
$\sim$5\,mas off the core (seen also in the RRFID images).

\acknowledgments

JAZ was supported for this research through 
a Max-Planck Research Award.
This research has made use of data from 
the University of Michigan Radio Astronomy
Observatory (supported by funds from the University of Michigan);
the United States Naval 
Observatory (USNO) Radio Reference Frame Image Database (RRFID);
the NASA/IPAC
Extragalactic Database (NED, operated by the Jet Propulsion Laboratory,
California Institute of Technology, under contract with the 
National Aeronautics and Space Administration); 
and the SIMBAD database (operated at CDS, Strasbourg, France).
H.\ M.\ Aller, 
M.\ C.\ Aller,
J.\ Armstrong, 
D.\ C.\ Homan, 
M.\ L.\ Lister, 
A.\ P.\ Lobanov 
M.\ Russo, 
and 
R.\ West, 
provided 
valuable 
help and important suggestions.
R.\ W.\ Goodrich and G.\ B.\ Taylor collaborated in the optical 
observations reported in the Appendix.

\clearpage

\appendix

\section{Three new spectroscopic redshifts\label{app:redshift}}

A 3000\,sec exposure on 0026+346 was obtained at the Palomar
200$^{\prime\prime}$ telescope by R.C.\ Vermeulen and G.B.\ Taylor on
1996 January 14, using the COSMIC spectrograph (\citealt{Kells98}).
with a 300\,g\,mm$^{-1}$ grism and a 1.5$^{\prime\prime}$ slit. 
Superimposed on a red continuum, the emission lines of 
[O{\sc ii}] $\lambda3727$ and
[O{\sc iii}] $\lambda4959$ are prominent, as is [O{\sc iii}]
$\lambda5007$, but the centroid of the latter is not well-determined
because it falls in the terrestrial atmospheric A-band absorption. From
the host galaxy, a Ca{\sc ii}\,H absorption line, 4000\,\AA-break, and
4300\,\AA\ G-band are also clearly visible. All features in the spectrum
are consistent with a redshift of $z=0.517\pm0.001$. 
In NED, 0026+346 is listed with a redshift of $z$=0.6.  As
far as we have been able to ascertain, this value
is a photometric estimate ---rather accurate, as we now find--- which
can be traced back uniquely to a paper by \citet{Hutchings94}.

Observations of 0727--115 and 1155+251 were made at the Keck~II
telescope by M.H.\ Cohen and R.W.\ Goodrich on 2000 January 8. 
The LRIS spectrograph \citep{Oke95} was used with the
polarimetry module \citep{Goodrich95} and the 300\,g/mm 
grating with the 1$^{\prime\prime}$ slit. The sky
had thin variable clouds.  

One exposure of duration 300\,sec was obtained for 0727--115.  This
object is only 3 degrees from the galactic plane and is faint, but
clearly extended.  Comparison with HD245310 ($V$=9.10; \citealt{Schmidt92})
yields $V \approx 20.3$.  
C{\sc iv} $\lambda1549$, C{\sc iii}] $\lambda1909$ and 
Mg{\sc ii} $\lambda2798$
are prominent and give z=1.5885, 1.5895, and 1.600 respectively.  The
C{\sc iii}] line is cleaner and has higher S/N than the others, and the
weighted redshift is 1.591$\pm0.003$.

There apparently is no published spectrum for 1155+251.
The Keck observations show that this
bright, narrow-line radio galaxy is reddened and has weak broad H$\alpha$
emission. The redshift is 0.2016$\pm0.0004$, based on a direct comparison between
the strong lines [O{\sc iii}] $\lambda\lambda4959,5007$ and the nearby night
sky lines [O{\sc i}] $\lambda5577$ and Na{\sc i} $\lambda5981$.  
Other prominent lines include [O{\sc ii}] $\lambda3727$, 
[Ne{\sc iii}] $\lambda3869$, H$\beta$,
H$\alpha$/[N{\sc ii}], and [S{\sc ii}] $\lambda\lambda6716,6730$.

\clearpage

\bibliography{Zensus}

\begin{thebibliography}{113}
\expandafter\ifx\csname natexlab\endcsname\relax\def\natexlab#1{#1}\fi

\bibitem[{{Aaron} {et~al.}(1997){Aaron}, {Wardle}, \& {Roberts}}]{Aaron97}
{Aaron}, S.~E., {Wardle}, J. F.~C., \& {Roberts}, D.~H. 1997, Vistas in
  Astronomy, 41, 225

\bibitem[{{Aaron} {et~al.}(1998){Aaron}, {Wardle}, \& {Roberts}}]{Aaron98}
{Aaron}, S.~E., {Wardle}, J. F.~C., \& {Roberts}, D.~H. 1998, in ASP Conf. Ser.
  144: IAU Colloq. 164, Radio Emission from Galactic and Extragalactic Compact
  Sources, ed. J.A. Zensus, G.B. Taylor, \& J.M. Wrobel (San Francisco: ASP),
  105

\bibitem[{{Aller} {et~al.}(1992){Aller}, {Aller}, \& {Hughes}}]{Aller92}
{Aller}, M.~F., {Aller}, H.~D., \& {Hughes}, P.~A. 1992, \apj, 399, 16

\bibitem[{{Axon} {et~al.}(2000){Axon}, {Capetti}, {Fanti}, {Morganti},
  {Robinson}, \& {Spencer}}]{Axon00}
{Axon}, D.~J., {Capetti}, A., {Fanti}, R., {Morganti}, R., {Robinson}, A., \&
  {Spencer}, R. 2000, \aj, 120, 2284

\bibitem[{{Beasley} {et~al.}(1996){Beasley}, {Dhawan}, {Fomalont}, \& {et
  al.}}]{Beasley96}
{Beasley}, A.~J., {Dhawan}, V., {Fomalont}, E.~B., \& {et al.} 1996, in IAU
  Symp. 175: Extragalactic Radio Sources, ed. R. Ekers, C. Fanti, \& L.
  Padrielli (Dordrecht: Kluwer), Vol. 175, 527

\bibitem[{{Bond}(1971)}]{Bond71}
{Bond}, H.~E. 1971, \apjl, 167, L79

\bibitem[{{Bondi} {et~al.}(1996){Bondi}, {Padrielli}, {Fanti}, {Ficarra},
  {Gregorini}, {Mantovani}, {Bartel}, {Romney}, {Nicolson}, \&
  {Weiler}}]{Bondi96}
{Bondi}, M., {Padrielli}, L., {Fanti}, R., {Ficarra}, A., {Gregorini}, L.,
  {Mantovani}, F., {Bartel}, N., {Romney}, J.~D., {Nicolson}, G.~D., \&
  {Weiler}, K.~W. 1996, \aap, 308, 415

\bibitem[{{Burbidge} {et~al.}(1974){Burbidge}, {Jones}, \&
  {Odell}}]{Burbidge74}
{Burbidge}, G.~R., {Jones}, T.~W., \& {Odell}, S.~L. 1974, \apj, 193, 43

\bibitem[{{Carilli} {et~al.}(1997){Carilli}, {Menten}, {Reid}, \&
  {Rupen}}]{Carilli97}
{Carilli}, C.~L., {Menten}, K.~M., {Reid}, M.~J., \& {Rupen}, M.~P. 1997,
  \apjl, 474, L89

\bibitem[{{Carini} {et~al.}(1991){Carini}, {Miller}, {Noble}, \&
  {Sadun}}]{Carini91}
{Carini}, M.~T., {Miller}, H.~R., {Noble}, J.~C., \& {Sadun}, A.~C. 1991, \aj,
  101, 1196

\bibitem[{{Cassaro} {et~al.}(1999){Cassaro}, {Stanghellini}, {Bondi},
  {Dallacasa}, {della Ceca}, \& {Zappal{\`a}}}]{Cassaro99}
{Cassaro}, P., {Stanghellini}, C., {Bondi}, M., {Dallacasa}, D., {della Ceca},
  R., \& {Zappal{\`a}}, R.~A. 1999, \aaps, 139, 601

\bibitem[{{Dallacasa} {et~al.}(1998){Dallacasa}, {Bondi}, {Alef}, \&
  {Mantovani}}]{Dallacasa98}
{Dallacasa}, D., {Bondi}, M., {Alef}, W., \& {Mantovani}, F. 1998, \aaps, 129,
  219

\bibitem[{{de Bruyn} {et~al.}(1996){de Bruyn}, {O'Dea}, \& {Baum}}]{DeBruyn96}
{de Bruyn}, A.~G., {O'Dea}, C.~P., \& {Baum}, S.~A. 1996, \aap, 305, 450

\bibitem[{{Djorgovski} {et~al.}(1987){Djorgovski}, {Perley}, {Meylan}, \&
  {McCarthy}}]{Djorgovski87}
{Djorgovski}, S., {Perley}, R., {Meylan}, G., \& {McCarthy}, P. 1987, \apjl,
  321, L17

\bibitem[{{Drinkwater} {et~al.}(1997){Drinkwater}, {Webster}, {Francis},
  {Condon}, {Ellison}, {Jauncey}, {Lovell}, {Peterson}, \&
  {Savage}}]{Drinkwater97}
{Drinkwater}, M.~J., {Webster}, R.~L., {Francis}, P.~J., {Condon}, J.~J.,
  {Ellison}, S.~L., {Jauncey}, D.~L., {Lovell}, J., {Peterson}, B.~A., \&
  {Savage}, A. 1997, \mnras, 284, 85

\bibitem[{{Ellison} {et~al.}(2001){Ellison}, {Pettini}, {Steidel}, \&
  {Shapley}}]{Ellison01}
{Ellison}, S.~L., {Pettini}, M., {Steidel}, C.~C., \& {Shapley}, A.~E. 2001,
  \apj, 549, 770

\bibitem[{{Fanti} {et~al.}(1995){Fanti}, {Fanti}, {Dallacasa}, {Schilizzi},
  {Spencer}, \& {Stanghellini}}]{Fanti95}
{Fanti}, C., {Fanti}, R., {Dallacasa}, D., {Schilizzi}, R.~T., {Spencer},
  R.~E., \& {Stanghellini}, C. 1995, \aap, 302, 317

\bibitem[{{Fey} \& {Charlot}(1997)}]{Fey97}
{Fey}, A.~L. \& {Charlot}, P. 1997, \apjs, 111, 95

\bibitem[{{Fey} \& {Charlot}(2000)}]{Fey00}
---. 2000, \apjs, 128, 17

\bibitem[{{Fey} {et~al.}(1996){Fey}, {Clegg}, \& {Fomalont}}]{Fey96}
{Fey}, A.~L., {Clegg}, A.~W., \& {Fomalont}, E.~B. 1996, \apjs, 105, 299

\bibitem[{{Fomalont} {et~al.}(2000){Fomalont}, {Frey}, {Paragi}, {Gurvits},
  {Scott}, {Taylor}, {Edwards}, \& {Hirabayashi}}]{Fomalont00}
{Fomalont}, E.~B., {Frey}, S., {Paragi}, Z., {Gurvits}, L.~I., {Scott}, W.~K.,
  {Taylor}, A.~R., {Edwards}, P.~G., \& {Hirabayashi}, H. 2000, \apjs, 131, 95

\bibitem[{{Gabuzda} {et~al.}(1999){Gabuzda}, {Pushkarev}, \&
  {Cawthorne}}]{Gabuzda99}
{Gabuzda}, D.~C., {Pushkarev}, A.~B., \& {Cawthorne}, T.~V. 1999, \mnras, 307,
  725

\bibitem[{{Gilmore} \& {Shaw}(1986)}]{Gilmore86}
{Gilmore}, G. \& {Shaw}, M.~A. 1986, \nat, 321, 750

\bibitem[{{Goodrich} {et~al.}(1995){Goodrich}, {Cohen}, \&
  {Putney}}]{Goodrich95}
{Goodrich}, R.~W., {Cohen}, M.~H., \& {Putney}, A. 1995, \pasp, 107, 179

\bibitem[{{Hardcastle} {et~al.}(1997){Hardcastle}, {Alexander}, {Pooley}, \&
  {Riley}}]{Hardcastle97}
{Hardcastle}, M.~J., {Alexander}, P., {Pooley}, G.~G., \& {Riley}, J.~M. 1997,
  \mnras, 288, L1

\bibitem[{{Hartman} {et~al.}(1999){Hartman}, {Bertsch}, {Bloom}, {Chen},
  {Deines-Jones}, {Esposito}, {Fichtel}, {Friedlander}, {Hunter}, {McDonald},
  {Sreekumar}, {Thompson}, {Jones}, {Lin}, {Michelson}, {Nolan}, {Tompkins},
  {Kanbach}, {Mayer-Hasselwander}, {M{\" u}cke}, {Pohl}, {Reimer}, {Kniffen},
  {Schneid}, {von Montigny}, {Mukherjee}, \& {Dingus}}]{Hartman99}
{Hartman}, R.~C., {Bertsch}, D.~L., {Bloom}, S.~D., {Chen}, A.~W.,
  {Deines-Jones}, P., {Esposito}, J.~A., {Fichtel}, C.~E., {Friedlander},
  D.~P., {Hunter}, S.~D., {McDonald}, L.~M., {Sreekumar}, P., {Thompson},
  D.~J., {Jones}, B.~B., {Lin}, Y.~C., {Michelson}, P.~F., {Nolan}, P.~L.,
  {Tompkins}, W.~F., {Kanbach}, G., {Mayer-Hasselwander}, H.~A., {M{\" u}cke},
  A., {Pohl}, M., {Reimer}, O., {Kniffen}, D.~A., {Schneid}, E.~J., {von
  Montigny}, C., {Mukherjee}, R., \& {Dingus}, B.~L. 1999, \apjs, 123, 79

\bibitem[{{Hewett} {et~al.}(1995){Hewett}, {Foltz}, \& {Chaffee}}]{Hewett95}
{Hewett}, P.~C., {Foltz}, C.~B., \& {Chaffee}, F.~H. 1995, \aj, 109, 1498

\bibitem[{{Hewitt} \& {Burbidge}(1989)}]{Hewitt89}
{Hewitt}, A. \& {Burbidge}, G. 1989, \apjs, 69, 1

\bibitem[{{Hughes} {et~al.}(1992){Hughes}, {Aller}, \& {Aller}}]{hughes92}
{Hughes}, P.~A., {Aller}, H.~D., \& {Aller}, M.~F. 1992, \apj, 396, 469

\bibitem[{{Hummel} {et~al.}(1992){Hummel}, {Muxlow}, {Krichbaum},
  {Quirrenbach}, {Schalinski}, {Witzel}, \& {Johnston}}]{Hummel92}
{Hummel}, C.~A., {Muxlow}, T. W.~B., {Krichbaum}, T.~P., {Quirrenbach}, A.,
  {Schalinski}, C.~J., {Witzel}, A., \& {Johnston}, K.~J. 1992, \aap, 266, 93

\bibitem[{{Hutchings} {et~al.}(1998){Hutchings}, {Dewey}, {Chaytor},
  {Ryneveld}, {Gower}, \& {Ellingson}}]{Hutchings98}
{Hutchings}, J.~B., {Dewey}, A., {Chaytor}, D., {Ryneveld}, S., {Gower}, A.~C.,
  \& {Ellingson}, E. 1998, \pasp, 110, 111

\bibitem[{{Hutchings} {et~al.}(1994){Hutchings}, {Neff}, {Weadock}, {Roberts},
  {Ryneveld}, \& {Gower}}]{Hutchings94}
{Hutchings}, J.~B., {Neff}, S.~G., {Weadock}, J., {Roberts}, L., {Ryneveld},
  S., \& {Gower}, A.~C. 1994, \aj, 107, 471

\bibitem[{{Hutchison} {et~al.}(2001){Hutchison}, {Cawthorne}, \&
  {Gabuzda}}]{Hutchison01}
{Hutchison}, J.~M., {Cawthorne}, T.~V., \& {Gabuzda}, D.~C. 2001, \mnras, 321,
  525

\bibitem[{{Hutsemekers}(1998)}]{hutsemekers98}
{Hutsemekers}, D. 1998, \aap, 332, 410

\bibitem[{{Impey} \& {Tapia}(1988)}]{Impey88}
{Impey}, C.~D. \& {Tapia}, S. 1988, \apj, 333, 666

\bibitem[{{Kanekar} \& {Chengalur}(1997)}]{Kanekar97}
{Kanekar}, N. \& {Chengalur}, J.~N. 1997, \mnras, 292, 831

\bibitem[{{Kedziora-Chudczer} {et~al.}(1998){Kedziora-Chudczer}, {Jauncey},
  {Tzioumis}, {Reynolds}, {Wieringa}, {Nicholson}, \&
  {Bignall}}]{KedzioraChudczer98}
{Kedziora-Chudczer}, L., {Jauncey}, D., {Tzioumis}, A., {Reynolds}, J.,
  {Wieringa}, M., {Nicholson}, G., \& {Bignall}, H. 1998, \iaucirc, 7066, 2

\bibitem[{{Kedziora-Chudczer} {et~al.}(1996){Kedziora-Chudczer}, {Jauncey},
  {Wieringa}, {Reynolds}, {Tzioumis}, \& {Nicholson}}]{KedzioraChudczer96}
{Kedziora-Chudczer}, L., {Jauncey}, D., {Wieringa}, M., {Reynolds}, J.,
  {Tzioumis}, A., \& {Nicholson}, G. 1996, \iaucirc, 6418, 2

\bibitem[{{Kedziora-Chudczer} {et~al.}(1997){Kedziora-Chudczer}, {Jauncey},
  {Wieringa}, {Walker}, {Nicolson}, {Reynolds}, \&
  {Tzioumis}}]{KedzioraChudczer97}
{Kedziora-Chudczer}, L., {Jauncey}, D.~L., {Wieringa}, M.~H., {Walker}, M.~A.,
  {Nicolson}, G.~D., {Reynolds}, J.~E., \& {Tzioumis}, A.~K. 1997, \apjl, 490,
  L9

\bibitem[{{Kedziora-Chudczer} {et~al.}(2001){Kedziora-Chudczer}, {Jauncey},
  {Wieringa}, {Tzioumis}, \& {Reynolds}}]{KedzioraChudczer01}
{Kedziora-Chudczer}, L.~L., {Jauncey}, D.~L., {Wieringa}, M.~H., {Tzioumis},
  A.~K., \& {Reynolds}, J.~E. 2001, \mnras, 325, 1411

\bibitem[{{Keel}(1985)}]{Keel85}
{Keel}, W.~C. 1985, \aj, 90, 2207

\bibitem[{{Kellermann} {et~al.}(1971){Kellermann}, {Jauncey}, {Cohen},
  {Shaffer}, {Clark}, {Broderick}, {R{\"o}nn{\"a}ng}, {Rydbeck}, {Matveyenko},
  {Moiseyev}, {Vitkevitch}, {Cooper}, \& {Batchelor}}]{Kellermann71}
{Kellermann}, K.~I., {Jauncey}, D.~L., {Cohen}, M.~H., {Shaffer}, B.~B.,
  {Clark}, B.~G., {Broderick}, J., {R{\"o}nn{\"a}ng}, B., {Rydbeck}, O. E.~H.,
  {Matveyenko}, L., {Moiseyev}, I., {Vitkevitch}, V.~V., {Cooper}, B. F.~C., \&
  {Batchelor}, R. 1971, \apj, 169, 1

\bibitem[{{Kellermann} \& {Pauliny-Toth}(1969)}]{Kellermann69}
{Kellermann}, K.~I. \& {Pauliny-Toth}, I.~I.~K. 1969, \apjl, 155, L71

\bibitem[{{Kellermann} {et~al.}(1998){Kellermann}, {Vermeulen}, {Zensus}, \&
  {Cohen}}]{kellermann98}
{Kellermann}, K.~I., {Vermeulen}, R.~C., {Zensus}, J.~A., \& {Cohen}, M.~H.
  1998, \aj, 115, 1295

\bibitem[{{Kells} {et~al.}(1998){Kells}, {Dressler}, {Sivaramakrishnan},
  {Carr}, {Koch}, {Epps}, {Hilyard}, \& {Pardeilhan}}]{Kells98}
{Kells}, W., {Dressler}, A., {Sivaramakrishnan}, A., {Carr}, D., {Koch}, E.,
  {Epps}, H., {Hilyard}, D., \& {Pardeilhan}, G. 1998, \pasp, 110, 1487

\bibitem[{{Kochanek} {et~al.}(1999){Kochanek}, {Falco}, \&
  {Mu{\~n}oz}}]{Kochanek99}
{Kochanek}, C.~S., {Falco}, E.~E., \& {Mu{\~n}oz}, J.~A. 1999, \apj, 510, 590

\bibitem[{{Kollgaard} {et~al.}(1990){Kollgaard}, {Wardle}, \&
  {Roberts}}]{Kollgaard90}
{Kollgaard}, R.~I., {Wardle}, J. F.~C., \& {Roberts}, D.~H. 1990, \aj, 100,
  1057

\bibitem[{{K\"onigl}(1981)}]{Koenigl81}
{K\"onigl}, A. 1981, \apj, 243, 700

\bibitem[{{Kovalev} {et~al.}(1999){Kovalev}, {Nizhelsky}, {Kovalev}, {Berlin},
  {Zhekanis}, {Mingaliev}, \& {Bogdantsov}}]{Kovalev99}
{Kovalev}, Y.~Y., {Nizhelsky}, N.~A., {Kovalev}, Y.~A., {Berlin}, A.~B.,
  {Zhekanis}, G.~V., {Mingaliev}, M.~G., \& {Bogdantsov}, A.~V. 1999, \aaps,
  139, 545

\bibitem[{{Kraus} {et~al.}(1999){Kraus}, {Witzel}, {Krichbaum}, {Lobanov},
  {Peng}, \& {Ros}}]{Kraus99}
{Kraus}, A., {Witzel}, A., {Krichbaum}, T.~P., {Lobanov}, A.~P., {Peng}, B., \&
  {Ros}, E. 1999, \aap, 352, L107

\bibitem[{{Krichbaum} {et~al.}(1990){Krichbaum}, {Hummel}, {Quirrenbach},
  {Schalinski}, {Witzel}, {Johnson}, {Muxlow}, \& {Qian}}]{Krichbaum90}
{Krichbaum}, T.~P., {Hummel}, C.~A., {Quirrenbach}, A., {Schalinski}, C.~J.,
  {Witzel}, A., {Johnson}, K.~J., {Muxlow}, T. W.~B., \& {Qian}, S.~J. 1990,
  \aap, 230, 271

\bibitem[{{L{\" a}hteenm{\" a}ki} \& {Valtaoja}(1999)}]{Lahteenmaki99b}
{L{\" a}hteenm{\" a}ki}, A. \& {Valtaoja}, E. 1999, \apj, 521, 493

\bibitem[{{L{\" a}hteenm{\" a}ki} {et~al.}(1999){L{\" a}hteenm{\" a}ki},
  {Valtaoja}, \& {Wiik}}]{Lahteenmaki99a}
{L{\" a}hteenm{\" a}ki}, A., {Valtaoja}, E., \& {Wiik}, K. 1999, \apj, 511, 112

\bibitem[{{Lawrence} {et~al.}(1985){Lawrence}, {Readhead}, {Linfield}, {Payne},
  {Preston}, {Schilizzi}, {Porcas}, {Booth}, \& {Burke}}]{Lawrence85}
{Lawrence}, C.~R., {Readhead}, A. C.~S., {Linfield}, R.~P., {Payne}, D.~G.,
  {Preston}, R.~A., {Schilizzi}, R.~T., {Porcas}, R.~W., {Booth}, R.~S., \&
  {Burke}, B.~F. 1985, \apj, 296, 458

\bibitem[{{Lawrence} {et~al.}(1996){Lawrence}, {Zucker}, {Readhead}, {Unwin},
  {Pearson}, \& {Xu}}]{Lawrence96}
{Lawrence}, C.~R., {Zucker}, J.~R., {Readhead}, A. C.~S., {Unwin}, S.~C.,
  {Pearson}, T.~J., \& {Xu}, W. 1996, \apjs, 107, 541

\bibitem[{{Lister}(2001)}]{Lister01b}
{Lister}, M.~L. 2001, \apj, 562, 208

\bibitem[{{Lister} {et~al.}(1998){Lister}, {Marscher}, \& {Gear}}]{Lister98}
{Lister}, M.~L., {Marscher}, A.~P., \& {Gear}, W.~K. 1998, \apj, 504, 702

\bibitem[{{Lister} {et~al.}(2001){Lister}, {Tingay}, \& {Preston}}]{Lister01}
{Lister}, M.~L., {Tingay}, S.~J., \& {Preston}, R.~A. 2001, \apj, 554, 964

\bibitem[{{Lobanov} {et~al.}(1998){Lobanov}, {Krichbaum}, {Witzel}, {Kraus},
  {Zensus}, {Britzen}, {Otterbein}, {Hummel}, \& {Johnston}}]{Lobanov98}
{Lobanov}, A.~P., {Krichbaum}, T.~P., {Witzel}, A., {Kraus}, A., {Zensus},
  J.~A., {Britzen}, S., {Otterbein}, K., {Hummel}, C.~A., \& {Johnston}, K.
  1998, \aap, 340, L60

\bibitem[{{Machalski}(1998)}]{machalski98}
{Machalski}, J. 1998, \aaps, 128, 153

\bibitem[{{Macomb} {et~al.}(1999){Macomb}, {Gehrels}, \& {Shrader}}]{Macomb99}
{Macomb}, D.~J., {Gehrels}, N., \& {Shrader}, C.~R. 1999, \apj, 513, 652

\bibitem[{{Macquart} {et~al.}(2000){Macquart}, {Kedziora-Chudczer}, {Rayner},
  \& {Jauncey}}]{Macquart00}
{Macquart}, J.~., {Kedziora-Chudczer}, L., {Rayner}, D.~P., \& {Jauncey}, D.~L.
  2000, \apj, 538, 623

\bibitem[{{Malizia} {et~al.}(2000){Malizia}, {Bassani}, {Dean}, {McCollough},
  {Stephen}, {Zhang}, \& {Paciesas}}]{Malizia00}
{Malizia}, A., {Bassani}, L., {Dean}, A.~J., {McCollough}, M., {Stephen},
  J.~B., {Zhang}, S.~N., \& {Paciesas}, W.~S. 2000, \apj, 531, 642

\bibitem[{{Martel} {et~al.}(1999){Martel}, {Baum}, {Sparks}, {Wyckoff},
  {Biretta}, {Golombek}, {Macchetto}, {de Koff}, {McCarthy}, \&
  {Miley}}]{Martel99}
{Martel}, A.~., {Baum}, S.~A., {Sparks}, W.~B., {Wyckoff}, E., {Biretta},
  J.~A., {Golombek}, D., {Macchetto}, F.~D., {de Koff}, S., {McCarthy}, P.~J.,
  \& {Miley}, G.~K. 1999, \apjs, 122, 81

\bibitem[{{Marziani} {et~al.}(1996){Marziani}, {Sulentic}, {Dultzin-Hacyan},
  {Calvani}, \& {Moles}}]{Marziani96}
{Marziani}, P., {Sulentic}, J.~W., {Dultzin-Hacyan}, D., {Calvani}, M., \&
  {Moles}, M. 1996, \apjs, 104, 37+

\bibitem[{Mattox {et~al.}(1997)Mattox, Schachter, Molnar, Hartman, \&
  Patnaik}]{mat97}
Mattox, J., Schachter, J., Molnar, L., Hartman, R., \& Patnaik, A. 1997, APJ,
  481, 95

\bibitem[{{Mattox} {et~al.}(1993){Mattox}, {Bertsch}, {Chiang}, {Dingus},
  {Fichtel}, {Hartman}, {Hunter}, {Kanbach}, {Kniffen}, {Kwok}, {Lin},
  {Mayer-Hasselwander}, {Michelson}, {von Montigny}, {Nolan}, {Pinkau},
  {Schneid}, {Sreekumar}, \& {Thompson}}]{Mattox93}
{Mattox}, J.~R., {Bertsch}, D.~L., {Chiang}, J., {Dingus}, B.~L., {Fichtel},
  C.~E., {Hartman}, R.~C., {Hunter}, S.~D., {Kanbach}, G., {Kniffen}, D.~A.,
  {Kwok}, P.~W., {Lin}, Y.~C., {Mayer-Hasselwander}, H.~A., {Michelson}, P.~F.,
  {von Montigny}, C., {Nolan}, P.~L., {Pinkau}, K., {Schneid}, E., {Sreekumar},
  P., \& {Thompson}, D.~J. 1993, \apj, 410, 609

\bibitem[{{Moellenbrock} {et~al.}(1996){Moellenbrock}, {Fujisawa}, {Preston},
  {Gurvits}, {Dewey}, {Hirabayashi}, {Inoue}, {Kameno}, {Kawaguchi}, {Iwata},
  {Jauncey}, {Migenes}, {Roberts}, {Schilizzi}, \& {Tingay}}]{Moellenbrock96}
{Moellenbrock}, G.~A., {Fujisawa}, K., {Preston}, R.~A., {Gurvits}, L.~I.,
  {Dewey}, R.~J., {Hirabayashi}, H., {Inoue}, M., {Kameno}, S., {Kawaguchi},
  M., {Iwata}, T., {Jauncey}, D.~L., {Migenes}, V., {Roberts}, D.~H.,
  {Schilizzi}, R.~T., \& {Tingay}, S.~J. 1996, \aj, 111, 2174

\bibitem[{{Morganti} {et~al.}(1993){Morganti}, {Killeen}, \&
  {Tadhunter}}]{Morganti93}
{Morganti}, R., {Killeen}, N. E.~B., \& {Tadhunter}, C.~N. 1993, \mnras, 263,
  1023

\bibitem[{{Murphy} {et~al.}(1993){Murphy}, {Browne}, \& {Perley}}]{Murphy93}
{Murphy}, D.~W., {Browne}, I. W.~A., \& {Perley}, R.~A. 1993, \mnras, 264, 298

\bibitem[{{Napier}(1994)}]{Napier94}
{Napier}, P.~J. 1994, in IAU Symp. 158: Very High Angular Resolution Imaging,
  ed. J.G. Robertson \& W.J. Tango (Dordrecht: Kluwer), Vol. 158, 117

\bibitem[{{Nilsson}(1998)}]{nilsson98}
{Nilsson}, K. 1998, \aaps, 132, 31

\bibitem[{{Oke} {et~al.}(1995){Oke}, {Cohen}, {Carr}, {Cromer}, {Dingizian},
  {Harris}, {Labrecque}, {Lucinio}, {Schaal}, {Epps}, \& {Miller}}]{Oke95}
{Oke}, J.~B., {Cohen}, J.~G., {Carr}, M., {Cromer}, J., {Dingizian}, A.,
  {Harris}, F.~H., {Labrecque}, S., {Lucinio}, R., {Schaal}, W., {Epps}, H., \&
  {Miller}, J. 1995, \pasp, 107, 375

\bibitem[{{Otterbein} {et~al.}(1998){Otterbein}, {Krichbaum}, {Kraus},
  {Lobanov}, {Witzel}, {Wagner}, \& {Zensus}}]{Otterbein98}
{Otterbein}, K., {Krichbaum}, T.~P., {Kraus}, A., {Lobanov}, A.~P., {Witzel},
  A., {Wagner}, S.~J., \& {Zensus}, J.~A. 1998, \aap, 334, 489

\bibitem[{{Owsianik} {et~al.}(1998){Owsianik}, {Conway}, \&
  {Polatidis}}]{Owsianik98}
{Owsianik}, I., {Conway}, J.~E., \& {Polatidis}, A.~G. 1998, \aap, 336, L37

\bibitem[{{Oya} {et~al.}(1998){Oya}, {Iwamuro}, {Tsukamoto}, \&
  {Maihara}}]{Oya98}
{Oya}, S., {Iwamuro}, F., {Tsukamoto}, H., \& {Maihara}, T. 1998, \pasj, 50,
  163

\bibitem[{{Patnaik} {et~al.}(1993){Patnaik}, {Browne}, {King}, {Muxlow},
  {Walsh}, \& {Wilkinson}}]{Patnaik93}
{Patnaik}, A.~R., {Browne}, I.~W.~A., {King}, L.~J., {Muxlow}, T.~W.~B.,
  {Walsh}, D., \& {Wilkinson}, P.~N. 1993, \mnras, 261, 435

\bibitem[{{Pearson} \& {Readhead}(1988)}]{Pearson88}
{Pearson}, T.~J. \& {Readhead}, A. C.~S. 1988, \apj, 328, 114

\bibitem[{{Pearson} {et~al.}(1994){Pearson}, {Shepherd}, {Taylor}, \&
  {Myers}}]{Pearson94}
{Pearson}, T.~J., {Shepherd}, M.~C., {Taylor}, G.~B., \& {Myers}, S.~T. 1994,
  American Astronomical Society Meeting, 185

\bibitem[{{Peng} {et~al.}(2000){Peng}, {Kraus}, {Krichbaum}, \&
  {Witzel}}]{Peng00}
{Peng}, B., {Kraus}, A., {Krichbaum}, T.~P., \& {Witzel}, A. 2000, \aaps, 145,
  1

\bibitem[{{Perley}(1982)}]{Perley82}
{Perley}, R.~A. 1982, \aj, 87, 859

\bibitem[{{Perlman} {et~al.}(1998){Perlman}, {Padovani}, {Giommi}, {Sambruna},
  {Jones}, {Tzioumis}, \& {Reynolds}}]{Perlman98}
{Perlman}, E.~S., {Padovani}, P., {Giommi}, P., {Sambruna}, R., {Jones}, L.~R.,
  {Tzioumis}, A., \& {Reynolds}, J. 1998, \aj, 115, 1253

\bibitem[{{Polatidis} \& {Wilkinson}(1998)}]{Polatidis98}
{Polatidis}, A.~G. \& {Wilkinson}, P.~N. 1998, \mnras, 294, 327

\bibitem[{{Polatidis} {et~al.}(1995){Polatidis}, {Wilkinson}, {Xu}, {Readhead},
  {Pearson}, {Taylor}, \& {Vermeulen}}]{Polatidis95}
{Polatidis}, A.~G., {Wilkinson}, P.~N., {Xu}, W., {Readhead}, A. C.~S.,
  {Pearson}, T.~J., {Taylor}, G.~B., \& {Vermeulen}, R.~C. 1995, \apjs, 98, 1

\bibitem[{{Pravdo} \& {Marshall}(1984)}]{Pravdo84}
{Pravdo}, S.~H. \& {Marshall}, F.~E. 1984, \apj, 281, 570

\bibitem[{{Price} {et~al.}(1993){Price}, {Gower}, {Hutchings}, {Talon},
  {Duncan}, \& {Ross}}]{Price93}
{Price}, R., {Gower}, A.~C., {Hutchings}, J.~B., {Talon}, S., {Duncan}, D., \&
  {Ross}, G. 1993, \apjs, 86, 365

\bibitem[{{Quirrenbach} {et~al.}(2000){Quirrenbach}, {Kraus}, {Witzel},
  {Zensus}, {Peng}, {Risse}, {Krichbaum}, {Wegner}, \&
  {Naundorf}}]{Quirrenbach00}
{Quirrenbach}, A., {Kraus}, A., {Witzel}, A., {Zensus}, J.~A., {Peng}, B.,
  {Risse}, M., {Krichbaum}, T.~P., {Wegner}, R., \& {Naundorf}, C.~E. 2000,
  \aaps, 141, 221

\bibitem[{{Rayner} {et~al.}(2000){Rayner}, {Norris}, \& {Sault}}]{Rayner00}
{Rayner}, D.~P., {Norris}, R.~P., \& {Sault}, R.~J. 2000, \mnras, 319, 484

\bibitem[{{Readhead}(1994)}]{Readhead94}
{Readhead}, A.~C.~S. 1994, \apj, 426, 51

\bibitem[{{Reid} {et~al.}(1995){Reid}, {Shone}, {Akujor}, {Browne}, {Murphy},
  {Pedelty}, {Rudnick}, \& {Walsh}}]{Reid95}
{Reid}, A., {Shone}, D.~L., {Akujor}, C.~E., {Browne}, I. W.~A., {Murphy},
  D.~W., {Pedelty}, J., {Rudnick}, L., \& {Walsh}, D. 1995, \aaps, 110, 213

\bibitem[{{Ros} {et~al.}(2001){Ros}, {Marcaide}, {Guirado}, \& {P{\'
  e}rez-Torres}}]{Ros01}
{Ros}, E., {Marcaide}, J.~M., {Guirado}, J.~C., \& {P{\' e}rez-Torres}, M.~A.
  2001, \aap, 376, 1090

\bibitem[{{Salgado} {et~al.}(1999){Salgado}, {Altschuler}, {Ghosh}, {Dennison},
  {Mitchell}, \& {Payne}}]{Salgado99}
{Salgado}, J.~., {Altschuler}, D.~R., {Ghosh}, T., {Dennison}, B.~K.,
  {Mitchell}, K.~J., \& {Payne}, H.~E. 1999, \apjs, 120, 77

\bibitem[{{Scarpa} {et~al.}(2000){Scarpa}, {Urry}, {Falomo}, {Pesce}, \&
  {Treves}}]{Scarpa00}
{Scarpa}, R., {Urry}, C.~M., {Falomo}, R., {Pesce}, J.~E., \& {Treves}, A.
  2000, \apj, 532, 740

\bibitem[{{Schmidt} {et~al.}(1992){Schmidt}, {Elston}, \& {Lupie}}]{Schmidt92}
{Schmidt}, G.~D., {Elston}, R., \& {Lupie}, O.~L. 1992, \aj, 104, 1563

\bibitem[{{Shen} {et~al.}(1997){Shen}, {Wan}, {Moran}, {Jauncey}, {Reynolds},
  {Tzioumis}, {Gough}, {Ferris}, {Sinclair}, {Jiang}, {Hong}, {Liang}, {Costa},
  {Tingay}, {McCulloch}, {Lovell}, {King}, {Nicolson}, {Murphy}, {Meier}, {van
  Ommen}, {Edwards}, \& {White}}]{Shen97}
{Shen}, Z.~., {Wan}, T.~., {Moran}, J.~M., {Jauncey}, D.~L., {Reynolds}, J.~E.,
  {Tzioumis}, A.~K., {Gough}, R.~G., {Ferris}, R.~H., {Sinclair}, M.~W.,
  {Jiang}, D.~., {Hong}, X.~., {Liang}, S.~., {Costa}, M.~E., {Tingay}, S.~J.,
  {McCulloch}, P.~M., {Lovell}, J. E.~J., {King}, E.~A., {Nicolson}, G.~D.,
  {Murphy}, D.~W., {Meier}, D.~L., {van Ommen}, R.~D., {Edwards}, P.~G., \&
  {White}, G.~L. 1997, \aj, 114, 1999

\bibitem[{{Shen} {et~al.}(1998){Shen}, {Wan}, {Moran}, {Jauncey}, {Reynolds},
  {Tzioumis}, {Gough}, {Ferris}, {Sinclair}, {Jiang}, {Hong}, {Liang},
  {Edwards}, {Costa}, {Tingay}, {McCulloch}, {Lovell}, {King}, {Nicolson},
  {Murphy}, {Meier}, {van Ommen}, \& {White}}]{Shen98}
{Shen}, Z.~., {Wan}, T.~., {Moran}, J.~M., {Jauncey}, D.~L., {Reynolds}, J.~E.,
  {Tzioumis}, A.~K., {Gough}, R.~G., {Ferris}, R.~H., {Sinclair}, M.~W.,
  {Jiang}, D.~., {Hong}, X.~., {Liang}, S.~., {Edwards}, P.~G., {Costa}, M.~E.,
  {Tingay}, S.~J., {McCulloch}, P.~M., {Lovell}, J. E.~J., {King}, E.~A.,
  {Nicolson}, G.~D., {Murphy}, D.~W., {Meier}, D.~L., {van Ommen}, T.~D., \&
  {White}, G.~L. 1998, \aj, 115, 1357

\bibitem[{{Smith} {et~al.}(2000){Smith}, {Lucey}, {Hudson}, {Schlegel}, \&
  {Davies}}]{Smith00}
{Smith}, R.~J., {Lucey}, J.~R., {Hudson}, M.~J., {Schlegel}, D.~J., \&
  {Davies}, R.~L. 2000, \mnras, 313, 469

\bibitem[{{Stanghellini} {et~al.}(1997){Stanghellini}, {O'Dea}, {Baum},
  {Dallacasa}, {Fanti}, \& {Fanti}}]{Stanghellini97}
{Stanghellini}, C., {O'Dea}, C.~P., {Baum}, S.~A., {Dallacasa}, D., {Fanti},
  R., \& {Fanti}, C. 1997, \aap, 325, 943

\bibitem[{{Stecker} \& {Tsuruta}(1972)}]{Stecker72}
{Stecker}, F.~W. \& {Tsuruta}, S. 1972, \nat, 235, 8

\bibitem[{{Stickel} {et~al.}(1994){Stickel}, {Meisenheimer}, \&
  {Kuehr}}]{Stickel94}
{Stickel}, M., {Meisenheimer}, K., \& {Kuehr}, H. 1994, \aaps, 105, 211

\bibitem[{{Taylor}(1998)}]{Taylor98}
{Taylor}, G.~B. 1998, \apj, 506, 637

\bibitem[{{Tornikoski} {et~al.}(1999){Tornikoski}, {Tingay}, {M{\"u}cke},
  {Chen}, {Connaughton}, {Jauncey}, {Johnston-Hollitt}, {Kemp}, {King},
  {McGee}, {Rantakyr{\"o}}, {Rayner}, {Reimer}, \& {Tzioumis}}]{Tornikoski99}
{Tornikoski}, M., {Tingay}, S.~J., {M{\"u}cke}, A., {Chen}, A., {Connaughton},
  V., {Jauncey}, D.~L., {Johnston-Hollitt}, M., {Kemp}, J., {King}, E.~A.,
  {McGee}, P., {Rantakyr{\"o}}, F., {Rayner}, D., {Reimer}, O., \& {Tzioumis},
  A.~K. 1999, \aj, 118, 1161

\bibitem[{{Ulvestad} {et~al.}(1983){Ulvestad}, {Johnston}, \&
  {Weiler}}]{Ulvestad83}
{Ulvestad}, J.~S., {Johnston}, K.~J., \& {Weiler}, K.~W. 1983, \apj, 266, 18

\bibitem[{{Valtaoja} {et~al.}(1999){Valtaoja}, {L{\" a}hteenm{\" a}ki}, {Ter{\"
  a}sranta}, \& {Lainela}}]{Valtaoja99}
{Valtaoja}, E., {L{\" a}hteenm{\" a}ki}, A., {Ter{\" a}sranta}, H., \&
  {Lainela}, M. 1999, \apjs, 120, 95

\bibitem[{{van Breugel} {et~al.}(1984){van Breugel}, {Miley}, \&
  {Heckman}}]{VanBreugel84}
{van Breugel}, W., {Miley}, G., \& {Heckman}, T. 1984, \aj, 89, 5

\bibitem[{{van Breugel} {et~al.}(1992){van Breugel}, {Fanti}, {Fanti},
  {Stanghellini}, {Schilizzi}, \& {Spencer}}]{VanBreugel92}
{van Breugel}, W.~J.~M., {Fanti}, C., {Fanti}, R., {Stanghellini}, C.,
  {Schilizzi}, R.~T., \& {Spencer}, R.~E. 1992, \aap, 256, 56

\bibitem[{{Venugopal} {et~al.}(1985){Venugopal}, {Ananthakrishnan}, {Swarup},
  {Pynzar}, \& {Udaltsov}}]{Venugopal85}
{Venugopal}, V.~R., {Ananthakrishnan}, S., {Swarup}, G., {Pynzar}, A.~V., \&
  {Udaltsov}, V.~A. 1985, \mnras, 215, 685

\bibitem[{{von Linde} {et~al.}(1993){von Linde}, {Borgeest}, {Schramm},
  {Graser}, {Heidt}, {Hopp}, {Meisenheimer}, {Nieser}, {Steinle}, \&
  {Wagner}}]{VonLinde93}
{von Linde}, J., {Borgeest}, U., {Schramm}, K.~., {Graser}, U., {Heidt}, J.,
  {Hopp}, U., {Meisenheimer}, K., {Nieser}, L., {Steinle}, H., \& {Wagner}, S.
  1993, \aap, 267, L23

\bibitem[{{Wiklind} \& {Combes}(1996)}]{Wiklind96}
{Wiklind}, T. \& {Combes}, F. 1996, \aap, 315, 86

\bibitem[{{Wiklind} \& {Combes}(1997)}]{Wiklind97}
---. 1997, \aap, 328, 48

\bibitem[{{Zensus}(1997)}]{Zensus97}
{Zensus}, J.~A. 1997, \araa, 35, 607

\bibitem[{{Zensus} {et~al.}(1995){Zensus}, {Cohen}, \& {Unwin}}]{Zensus95}
{Zensus}, J.~A., {Cohen}, M.~H., \& {Unwin}, S.~C. 1995, \apj, 443, 35

\bibitem[{{Zensus} {et~al.}(1984){Zensus}, {Porcas}, \&
  {Pauliny-Toth}}]{Zensus84}
{Zensus}, J.~A., {Porcas}, R.~W., \& {Pauliny-Toth}, I. I.~K. 1984, \aap, 133,
  27

\end{thebibliography}

\end{document}